\documentclass[%
 reprint,
 amsmath,amssymb,
 aps,aip,graphicx,
]{revtex4-1}

\usepackage{graphicx}
\usepackage{dcolumn}
\usepackage{bm}

\usepackage{amsthm}
\usepackage{mathrsfs}
\usepackage{amssymb}
\usepackage{epsfig}
\usepackage{amsmath}
\usepackage{epsfig}
\usepackage{booktabs}
\usepackage{multirow}
\usepackage[colorlinks,urlcolor=blue]{hyperref}
\usepackage{makecell}
\usepackage{bm}
\usepackage[utf8]{inputenc}
\usepackage{color}
\begin{document}

\preprint{APS/123-QED}

\title{High-topological-number skyrmions and phase transition in two-dimensional frustrated $J_1$-$J_2$ magnets}

\author{Hongliang Hu$^{1}$\textsuperscript{\ddag}}

\author{Zhong Shen$^{1}$\textsuperscript{\ddag}}%

\author{Zheng Chen$^{1}$}
\author{Xiaoping Wu$^{1}$}
\author{Tingting Zhong$^{1}$}

\author{Changsheng Song$^{1,2}$}
\email{cssong@zstu.edu.cn}

\affiliation{$^{1}$Key Laboratory of Optical Field Manipulation of Zhejiang Province, Department of Physics, Zhejiang Sci-Tech University, Hangzhou 310018, China\\
$^{2}$Longgang Institute of Zhejiang Sci-Tech University, Wenzhou 325802, China}

\footnotetext{$\ddag~These authors contributed equally to this work.$}






\begin{abstract}
With the rapidly expanded field of two-dimensional(2D) magnetic materials, the frustrated magnetic skyrmions are attracting growing interest recently. Here, based on hexagonal close-packed (HCP) lattice of $J_1$-$J_2$ Heisenberg spins model, we systematically investigate the frustrated skyrmions and phase transition by micromagnetic simulations and first-principles calculations. The results show that four spin phases of antiferromagnetic, labyrinth domain, skyrmion and ferromagnetic textures are determined by the identified ranges of $J_1$-$J_2$. Importantly, skyrmion phase with an increasing topological number ($Q$) covers a wider $J_1$-$J_2$ area. Then, the diameter of skyrmions can be tuned by the frustration strength ($|J_2/J_1|$) or external magnetic field. Besides, a phase transition from N$\acute{e}$el to Bloch type skyrmion is observed due to the change of the helicity with the variation of $|J_2/J_1|$. Furthermore, as increasing magnetic field, the skyrmions with high $Q$ ($\ge 3$) tend to split into the ones with $Q=1$, thereby achieving a lower systematic energy. Additionally, we find that the CoCl$_2$ monolayer satisfies the requirement of the frustrated $J_1$-$J_2$ magnet, and the related magnetic behaviors agree with the above conclusions. The frustration-induced skyrmions are stable without the manipulation of temperature and magnetic field. Our results may open a possible way toward spintronic applications based on High-topological-number and nanoscale topological spin textures of skyrmions.
\end{abstract}

\maketitle


\section{\label{sec:level1}Introduction}

Skyrmions have been one of the most appealing fields of spintronics, since the experimental observation in MnSi\cite{RN144} and Fe$_{1-x}$Co$_x$Si\cite{RN156} with B20-type chiral crystal structure\cite{RN67}. Particularly, in recent years, the investigations of skyrmions have experienced rapidly and considerable development not only in the deep physics of skyrmions but also the application aspect such as storage\cite{RN5232,RN5233}, logic\cite{RN5234,RN5235,RN5236,RN5237} and neuromorphic\cite{RN5238,RN5239} devices. Most studies of skyrmions focus on the chiral magnets\cite{RN5293} with the Dzyaloshinskii-Moriya interaction (DMI)\cite{RN130,RN131}. The chirality of Janus monolayers\cite{RN130,RN131} with anisotropic DMI\cite{RN3742} is reported. Then, the chirality of non-coplanar spin structures, how chirality controls magnetic textures and offers insights into utilizing chiral magnets for spintronics through modulation of topological magnetic textures\cite{RN31,RN3688,RN1041}. However, the vorticity and helicity of skyrmions stabilized by DMI is fixed in these systems and the instability of skyrmions with  topological number ($Q$) greater than 2 is attributed to DMI's inability to prevent the skyrmion from collapsing into a point, ultimately resulting in the annihilation of the skyrmion\cite{RN199}, which leads to the scarcity of investigation on the phase transition of skyrmions with higher $Q$.

The frustrated magnets have been proven to host skyrmions both theoretically\cite{RN3326,RN3020,RN3728} and experimentally\cite{RN3103,RN152}. More importantly, the skyrmions in frustrated magnets possess the additional degrees of freedom\cite{RN116,RN5251,RN151,RN3741}, vorticity and helicity, making it possible to achieve the coexist of the skyrmions with different $Q$ and chiralities\cite{RN3326,RN153,RN105}, which is different from that of the DMI system. Generally, different ranges of frustration strength ($|J_2$/$J_1|$) are critical on the spin configuration, resulting in various phases including ferromagnetic, antiferromagnetic, skyrmion, labyrinth domain, helical state,  multisublattice helical, etc \cite{RN81,RN80,RN82,RN79,RN135}.  
Meanwhile, the spin frustration also has impact on skyrmion spin flipping, thus playing an important role in the regulation of topological-number of skyrmion\cite{RN84,RN85,RN83}. So far, the investigations of high-$Q$ skyrmions and their phase transition in frustrated systems are rather scarce, and it is of great importance to explore the collapse energy barriers of skyrmions as well as how the external magnetic field, magnetic anisotropy and frustration strength affect the topological number $Q$, vorticity, helicity and the diameter of skyrmions.

Here, the skyrmions with different $Q$ in frustrated 2D hexagonal close-packed (HCP) lattice are numerically investigated based on the spin Heisenberg model with the ferromagnetic nearest-neighbor(NN) $J_1$ and antiferromagnetic second NN exchange interactions $J_2$. The results show that the spin textures of ferromagnetic, labyrinth domain, skyrmion and antiferromagnetic phases are induced within the given ranges of $J_1$-$J_2$. Among them, the skyrmion phase covers a wider area with an increasing topological number Q. Besides, via varying the $|J_2$/$J_1|$, the skyrmion undergoes a phase transition between the N\'{e}el and the Bloch type, accompanied by a changed helicity. Moreover, the diameter of skyrmions reduces with the decreasing $|J_2$/$J_1|$ and the reducing rate is associated with the value of $Q$. Furthermore, the skyrmions with a higher $Q$ ($Q$$\geq$3) can be split into the ones with $Q=1$ and achieve a lower systematic energy as the increasing magnetic field. Particularly, based on density-functional theory (DFT) calculations, the CoCl$_2$ monolayer satisfies the frustrated  $J_1$-$J_2$ model,  and the induced small size ($<$10 nm) skyrmions are stable.

\section{\label{sec:level2}MODEL AND METHOD}

The spin system of frustrated HCP lattice is described by a classical Heisenberg $J_1$-$J_2$ model\cite{RN155,RN154} with the Hamiltonian being:
\begin{equation}\label{eq1}
\scriptsize
	H = - \sum\limits_{\left \langle i,j \right \rangle}J_{1}({\mathbf S_{i}}\cdot{\mathbf S_{j}}) - \sum\limits_{\left \langle k,l \right \rangle}J_{2}({\mathbf S_{k}}\cdot{\mathbf S_{l}}) - 
K\sum_{i}(\mathbf S_{i}^{z})^2 - \mu B\sum_{i}\mathbf S_{i}^{z}. 
\end{equation}
Here, $J_1$, $J_2$ are the Heisenberg exchange coefficients of the NN and second NN magnetic atoms with the positive (negative) values favoring ferromagnetic (antiferromagnetic) alignment, respectively. The single ion anisotropy and external magnetic field are characterized by the parameters $K$ and $B$, respectively. The $\mu$ is the magnetic moment of magnetic atom and ${\mathbf S}_{i (j,k,l)}$ representing the spin at site $i$ ($j,k ,l$).

Based on the above Hamiltonian, we perform micromagnetic simulations by varying frustration strength ($\lvert J_2/J_1 \rvert$) and external magnetic fields to explore the stability and structural details of skyrmions with high topological number $Q$ in 2D $J_1$-$J_2$ frustrated system. All micromagnetic simulations are performed using the framework of $Spirit$ package\cite{RN1055} with Landau-Lifshitz-Gilbert equation\cite{RN1984,RN1985}. In order to ensure that the energy reaches its minimum and the system achieves a stable state, all micromagnetic simulations were run for a total of 200,000 iteration steps as shown in Fig. S2 (in the Supplemental Material\cite{SM}). In addition, we use a large supercell containing 32400 magnetic atoms and consider periodic boundary conditions as shown in Fig. S3 (in the Supplemental Material\cite{SM}). The supercell size we adopted is 60 × 30 × 1, which is sufficiently large to prevent any distortions in the shape of skyrmions.  Besides, the geodesic nudged elastic band (GNEB)\cite{RN5280,RN5278} method is employed to calculate the minimum energy transition path and energy barrier of the skyrmion collapse.

Then we use the framework of DFT as implemented in the $Vienna$ $ab$ $initio$ $Simulation$ Package (VASP)\cite{RN70} to perform our first-principles calculations on the CoCl$_2$ monolayer, with the projected augmented-wave method\cite{RN71,RN73,RN72}describing the electron-core interaction. The generalized gradient approximation of Perdew-Burke-Ernzerhof\cite{RN74} is chosen to treat the exchange correlation effects with an effective Hubbardlike term U = 3 eV for 3$d$ electrons of Co\cite{RN145,RN146,RN147,RN148}. The plane-wave cutoff energy is set as 500 eV, and the first Brillouin-zone integration is carried out using 11 × 11 × 1 and 4 × 6 × 1 $\Gamma$-centered k-point meshes for the primitive cell and the 4 × 4 × 1 supercell, respectively. In order to accurately determine the exchange coefficient $J$, we set a high convergence standard with the energy and force less than $10^{-6}$ eV and 0.01eV/\AA, respectively.

\section{Results and discussion}
\begin{figure*}[t] 
	\begin{center}
		\includegraphics[width=14.0cm]{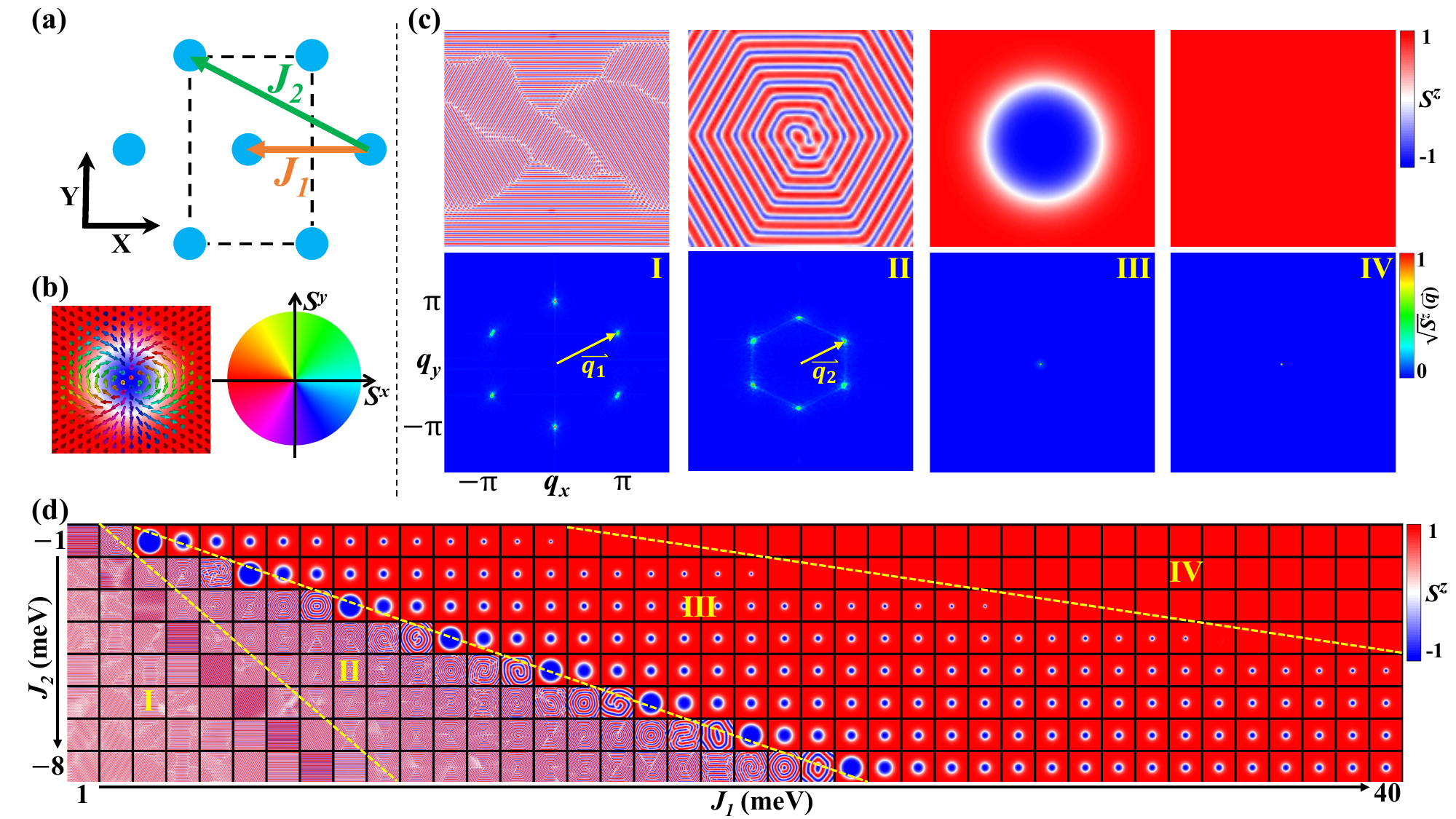}\
		\caption{\label{fig1}
			(a) Schematic of the HCP lattice, and the Heisenberg exchange interactions between the NN ($J_1$) and second NN ($J_2$) magnetic atoms. (b) The skyrmion with $Q = $ 2. The background shows the $z$ component of spins and the color of each spin is determined by their in-plane component as shown in the right channel of (b). (c) The four magnetic phases of AFM, labyrinth domain (LD), skyrmion and FM (top panel), and the square root of the z components of the spin structure factor (bottom panel). (d) The spin textures with varying $J_1$ and $J_2$, where $Q = $ 2.}
\end{center}
\end{figure*}
The basic geometric structure of HCP lattice is demonstrated in Fig. \ref{fig1}(a) where the orange and green arrows represent the NN and second NN Heisenberg exchange coefficients, respectively. In our micromagnetic simulations, we choose a skyrmion (Fig. \ref{fig1}(b)) with a high topological number of $Q$ =  2 in the center of the FM background as the initial state. Then, by varying the values of $J_1$ and $J_2$ in the range of 1 $\sim$ 40 meV and $-8 \sim -1$ meV, respectively, there are four phases induced as shown in the upper panel of Fig. \ref{fig1}(c), named antiferromagnet (AFM), labyrinth domain (LD), skyrmion, and ferromagnet (FM), respectively. And the LD state exhibits a resemblance to an Archimedean spiral, which has been elaborately described in the first section of the Supplemental Material\cite{SM}. The square root of the z components of spin structure factors are also demonstrated to characterize different phases as depicted in the lower panel of Fig. \ref{fig1}(c), which is convenient to compare with the diffraction peak image in experiment. Here, the spin structure factor is defined as\cite{RN3726,RN3724,RN2989}: $S^{\alpha }(\vec{q}) = \frac{1}{N} \sum\limits_{i,j }S^{\alpha}_{i}S^{\alpha}_{j}e^{i\vec{q} \cdot  (\vec{r}_i - \vec{r}_j)},$
where $\alpha = x, y, z$. The reciprocal spatial image obtained by Fourier-transforming of the spin texture reveals similarity in AFM and LD phases, each of which consists of six peaks forming a hexagon, attributed to the HCP lattice. However, the norm of the wave vector $\vec{q}_1$ is larger than that of $\vec{q}_2$, suggesting that the domain wall of AFM phase is narrower than in the LD phase.
 
Figure. \ref{fig1}(d) shows the magnetic phase diagram of AFM (\uppercase\expandafter{\romannumeral1}), LD (\uppercase\expandafter{\romannumeral2}), skyrmion (\uppercase\expandafter{\romannumeral3}) and FM (\uppercase\expandafter{\romannumeral4}) with varying $J_1$ and $J_2$. We notice that the boundary between area \uppercase\expandafter{\romannumeral2} and \uppercase\expandafter{\romannumeral3} is a straight line indicating that $|J_1| \geq  2.75|J_2|$ is required for the stability of skyrmion which is similar to that of the square lattice\cite{RN3719}. Moreover, the frustration is induced by competition between $J_1$ (FM) and $J_2$ (AFM). As $J_1$ increases, the frustration strength ($|J_2/J_1|$) decreases, resulting in a weaker competition between FM and AFM ordering. This gradual transition in the spin phase diagram of Fig. \ref{fig1}(d) leads to a transformation finally toward the FM state from AFM, LD and skyrmion. Meanwhile the diameter of skyrmions gradually shrinks with the increasing $J_1$ or the decreasing $J_2$  until skyrmions eventually annihilate into the FM ordering. Interestingly, as shown in Fig. S4 (in the Supplemental Material\cite{SM}), since the existence of the stronger topological protection in chirality skrmion, the larger the topological charge number, the wider the area where skyrmion exists.

\begin{figure}[!htbp]
	\begin{center}
		\centering
		\includegraphics[width=9.0cm]{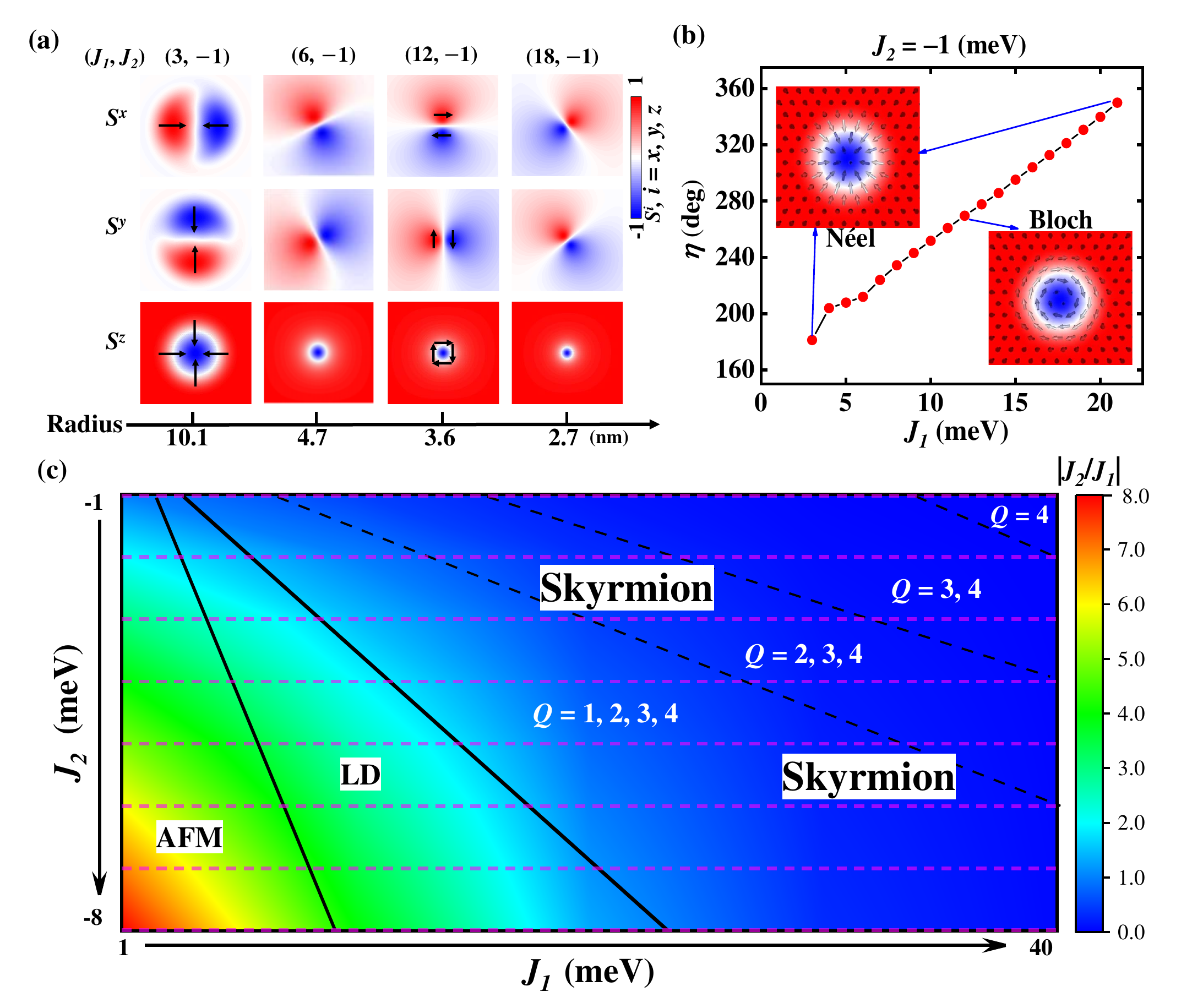}\
		\caption{\label{fig2}
			(a) The spin textures and radius of the skyrmions with a topological number of $Q=$ 1. (b) The helicity of skyrmions as a function of $J_1$. The insets of (b) show the N$\acute{e}$el and Bloch type skyrmions. (c) Phase diagram of the AFM, LD and skyrmion phase in the $J_1$-$J_2$ plane.
}
\end{center}
\end{figure}

Furthermore, in order to explore the structural characteristics of skyrmion under varying frustration strength ($|J_2/J_1|$), we visualize the spin textures with different spin components of skyrmion. Figure. \ref{fig2}(a) shows the spin components in $x$ (first line),  $y$ (second line) and $z$ (third line) directions. Herein, the frustration strength decreases with the increase of $J_1$ when the $J_2$ is fixed as $-1$ meV. It can be observed that the spin textures of $S^x$ and $S^y$ rotate around the center of the lattice as the increasing $J_1$. Then, the helicity ($\eta$) of skyrmion is described as the angle between the in-plane spin component and the normal component and it is defined by the phase factor $\eta$ in\cite{RN3688}: $\varPhi (\varphi ) = m\varphi + \eta.$ as shown in Fig. S5(b) (in the Supplemental Material\cite{SM}), where $\varPhi$ ($\varphi$) is the angle between X axis and the in-plane component of the spin (position) vector. $m=\pm 1$ is the vorticity which is related to the topological number of skyrmion. Interestingly, as illustrated in Fig. \ref{fig2}(b), the $\eta$ of skyrmion increases almost linearly from 180$^{\circ}$ to 270$^{\circ}$ and then to $\sim$ 360$^{\circ}$ as a function of $J_1$, suggesting that the skyrmion undergoes a  phase  transition  from  the Néel to Bloch type. In Néel skyrmions, spins rotate in a plane perpendicular to the core magnetization, while in Bloch skyrmions, spins rotate in a plane containing the core magnetization. The transition between these two types entails a change in the rotational sense of spins, with helicity serving as a quantitative measure of this rotation. And the topological configurations of Bloch and Néel types of skyrmions are described in the fourth section of the Supplemental Material\cite{SM}. Meanwhile, the generation and structural characteristics of skyrmions with different topological numbers ($Q =1, 2, 3, 4$) are shown in Fig. \ref{fig2}(c), which demonstrates the phase diagram of AFM, LD and skyrmion states as functions of $J_1$ and $J_2$.  It is noteworthy that the skyrmion with higher $Q$ (stronger topological protection) can exists a larger value of $J_1$-$J_2$, and it means that a greater energy is needed to annihilate the skyrmion with the higher $Q$.

\begin{figure}[h]
	\begin{center}
		\includegraphics[width=9.0cm]{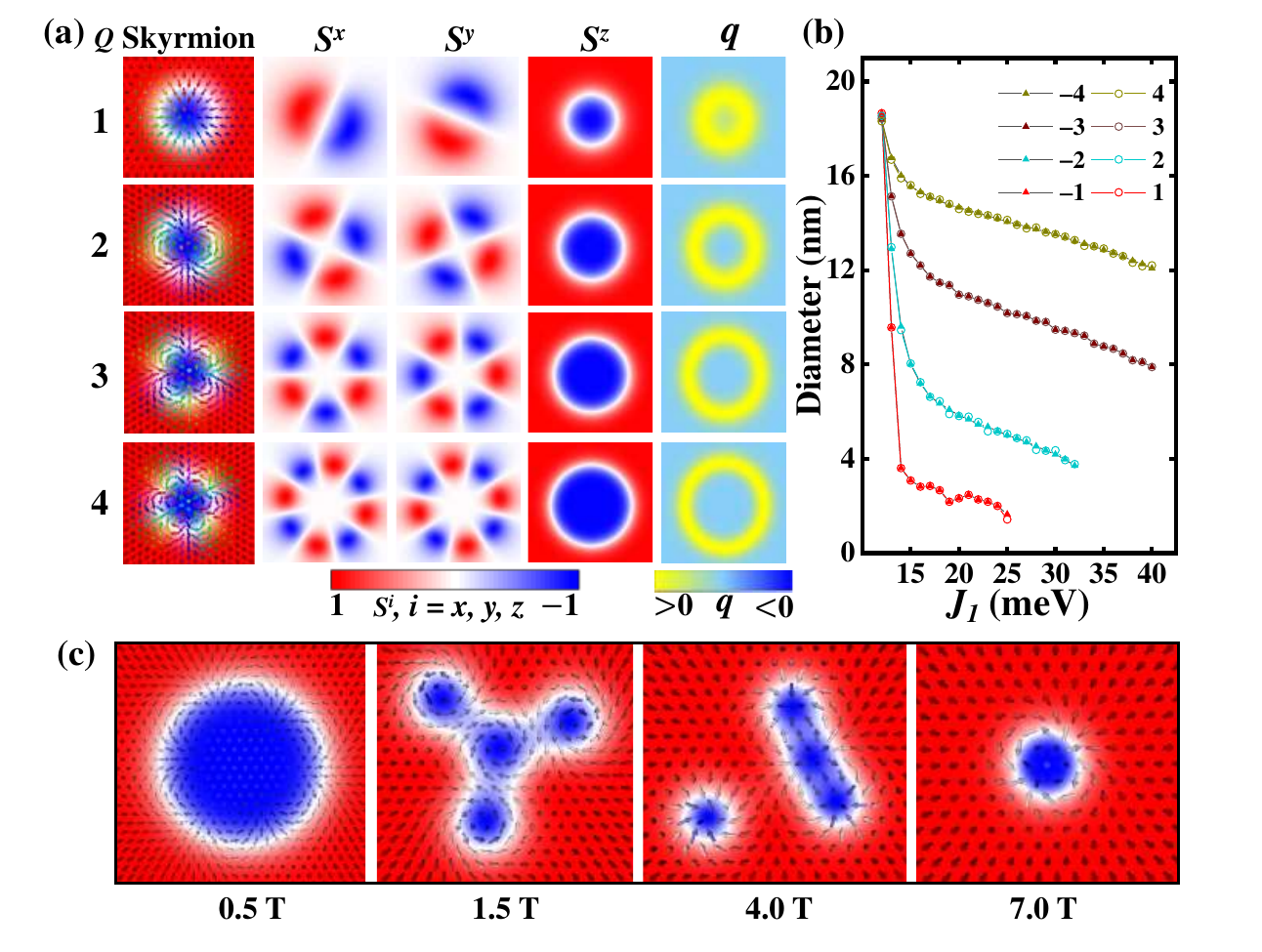}\
		\caption{\label{fig3}
			(a) The schematic of skyrmions with $Q = $ 1, 2, 3 and 4 (first column), the spin textures of the $x$ (second column), $y$ (third column) and $z$ (fourth column) components of spins as well as the topological charge density (last column). (b) The diameter of skyrmions and antiskyrmions with different topological number as functions of $J_1$ where $J_2$ is fixed at $-4$ meV. (c) The spin textures of skyrmion with $Q = $ 4 under different magnetic fields.
		}
	\end{center}
\end{figure}

The schematic diagrams of skyrmions with different $Q$ are shown in the first column of Fig. \ref{fig3}(a). Then, the 2 - 5 columns demonstrate the $x$, $y$, $z$ components of spin ($S^x$, $S^y$ and $S^z$) and topological charge density $q$ of skyrmion with $Q = $ 1, 2, 3 and 4, respectively. The number of red (blue) area in $S^x$ or $S^y$ textures equals to the topological number of skyrmion which counts how many times the spin rotates in $x$ $-$ $y$ plane. The $q$ of the last column in Fig. \ref{fig3}(a) shows that the topological charge density mainly distributed on the ring-shaped contour around the core of skyrmion. It's worth noting that the contour of $q$ is almost overlaps with the profile of $S^z = 0$ (the fourth column of Fig. \ref{fig3}(a)), which is mainly caused by the strong in-plane rotation of spins at the $S^z = 0$ contour. In addition, the diameter regulation of the skyrmion by frustration strength ($\lvert J_2/J_1 \rvert$, with a fixed $J_2$ of $-4$ meV) is investigated in Fig. \ref{fig3}(b), the diameter of skyrmion gradually decreases with the increased $J_1$ (the reduced frustration strength). Obviously, the reduction rate of skyrmion diameter is mainly related to the magnitude rather than the sign of $Q$, which means that the diameters of skyrmions and antiskyrmions undergo a similar trend. Moreover, we observed that, owing to the topological protection \cite{RN199}, the diameter of the skyrmion is less affected by $J_1$ with high-$Q$. Specifically, its diameter exhibits a slower decrease with the increase of $J_1$. Furthermore, the diameter demonstrates a linear decrease with the increase of $J_1$, when $J_1$ exceeds 17 meV. Furthermore, as shown in Fig. \ref{fig3}(c), the skyrmion with $Q = $ 4 maintains stable under a small magnetic field but splits into four skyrmions with $Q = 1$ in the external magnetic field of 1.5 T. The total topological number remains unchanged regardless of whether the splitting of the skyrmion begins or not even under a large magnetic field of 4.0 T. However, upon applying a larger magnetic field of 7.0 T, the topological protection of skyrmion is damaged resulting in the annihilation of the skyrmions with $Q = 1$, leaving only an isolated skyrmion with $Q = 2$, indicating that the magnetic field is very efficient at manipulating the topological number of skyrmions.

\begin{figure}[h]
	\begin{center}
	\includegraphics[width=9.0cm]{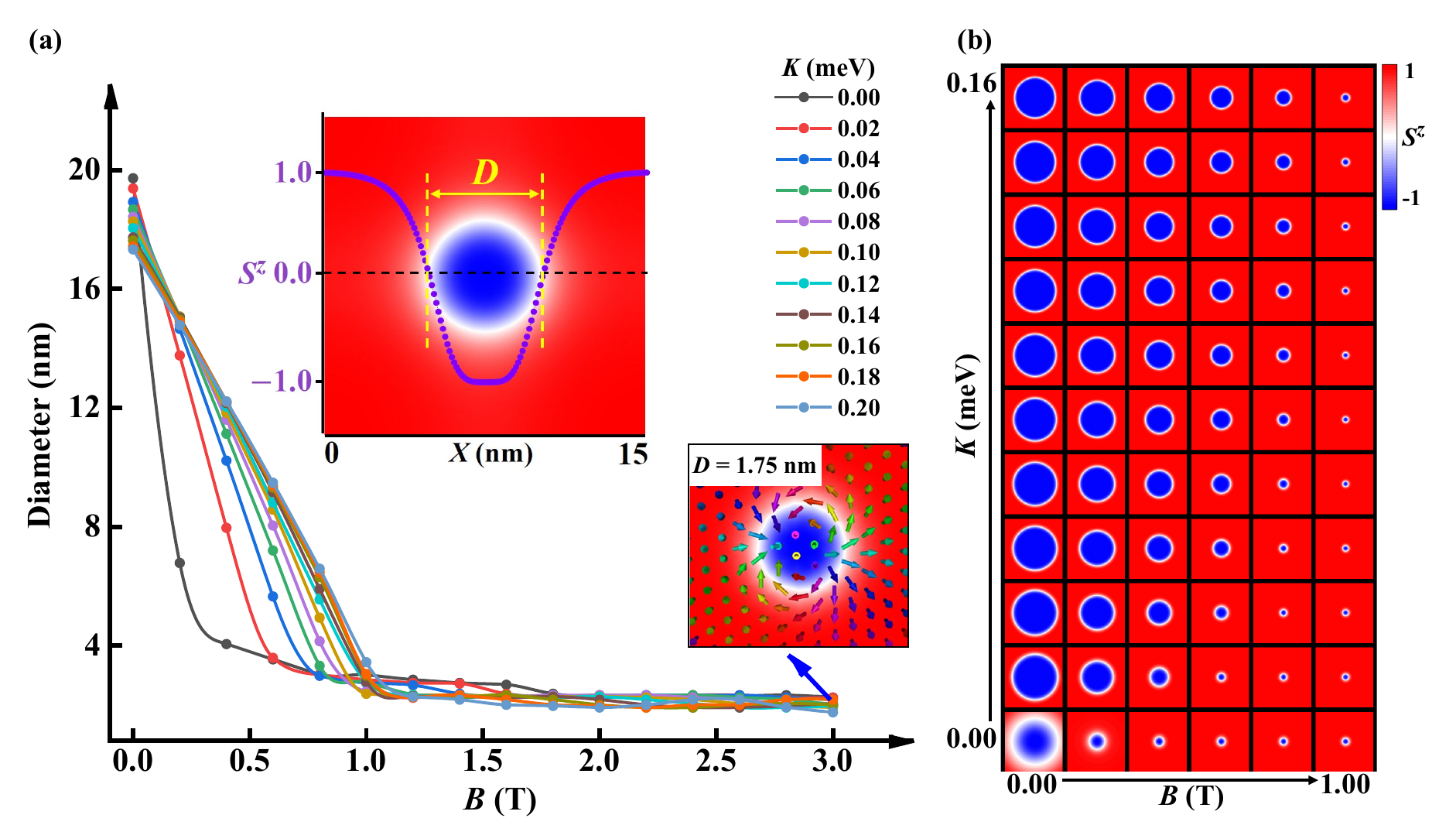}\
	\caption{\label{fig4}
	(a) The diameters of skyrmions as functions of external magnetic field ($B$) with different single ion anisotropy coefficients ($K$) and (b) corresponding spin textures with changing $B$ and $K$. The upper lift inset of (a) shows the diameter ($D$) of skyrmion which is defined as the diameter of the $S^z$ = 0 contour and the lower right inset demonstrates the minimum diameter of a skyrmion with $Q$ = 2.  The parameters used here are $J_1 = 3$ meV, $J_2 = -1$ meV and $Q = 2$.
	}
	\end{center}
	\end{figure}

The diameter of skyrmions can be manipulated not only via the frustration strength $|J_2/J_1|$ mentioned above, but also by the external magnetic field $B$ which has been proven to be a direct and effective way in many skyrmion systems\cite{RN3705,RN5231,RN31}. As shown in Fig. \ref{fig4}(a), the diameter of skyrmion significantly reduces to less than 4 nm with applying a small magnetic field. Furthermore, the skyrmion diameter is determined by the competition between the  external  magnetic  field $B$ and the single ion anisotropy $K$, The result of Fig. \ref{fig4}(b) illustrates that the size of skyrmion is more sensitive to magnetic field with a small $K$. However, it's important to note that, unlike in the continuous model, the diameter of the skyrmion cannot continuously shrink to zero within the discrete lattice. This explains why the skyrmions' diameter remains relatively unchanged within the range of $B$ from 1.0 T to 3.0 T. The inset in the lower right corner of Fig. \ref{fig4}(a) illustrates the minimum diameter of the skyrmion, which is approximately 1.75 nm.

\begin{figure}[!htbp]
	\begin{center}
		\includegraphics[width=9cm]{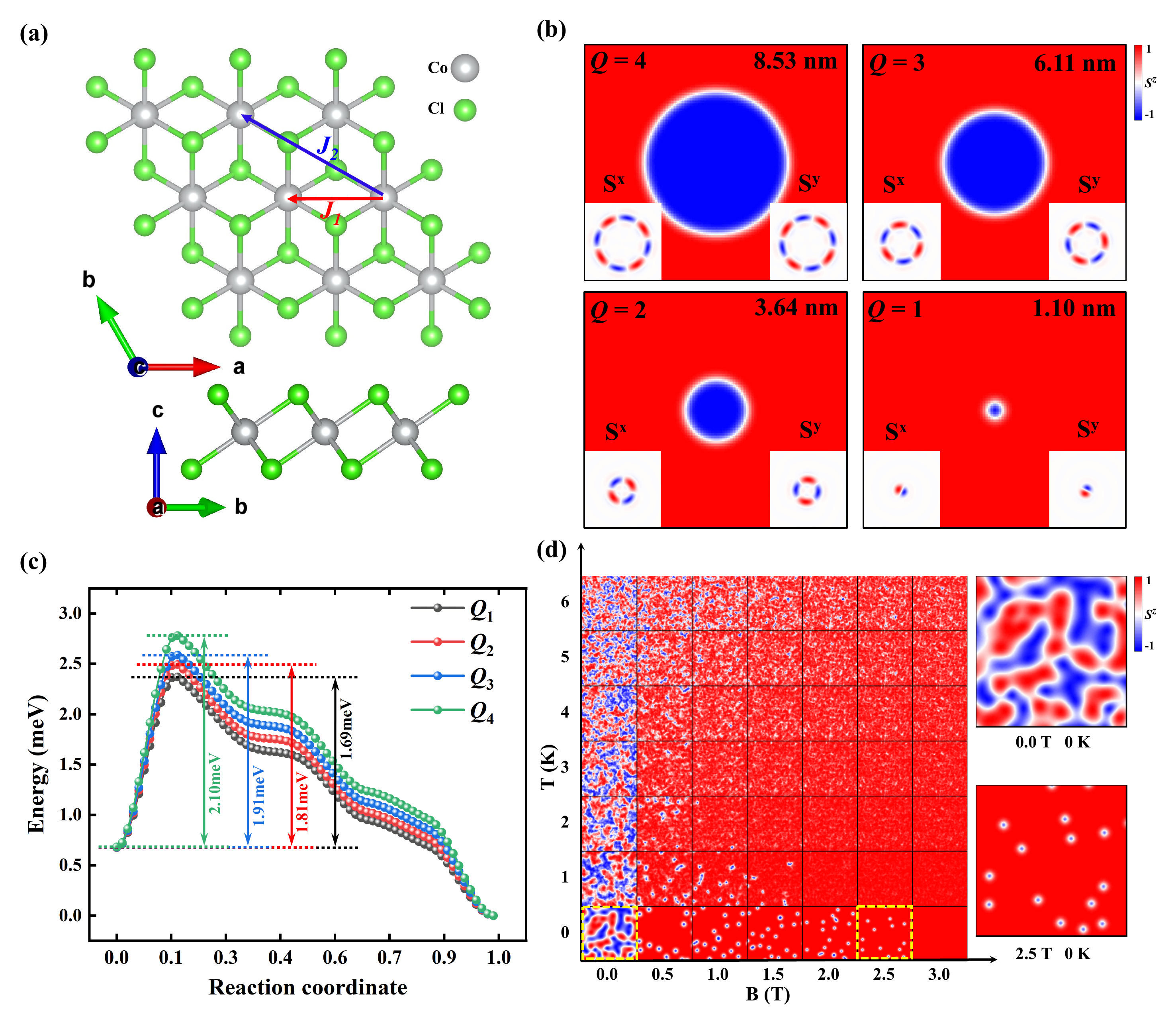}\
		\caption{\label{fig5}
		(a) Top and side views of CoCl$_2$ monolayer with the red and blue arrows being the Heisenberg exchange coefficients between the NN and second NN magnetic atoms. (b) The skyrmions with $Q = 4, 3, 2, 1$ and the insets in each subgraph show the spin textures of $S^x$ and $S^y$ under a small external magnetic field of 0.1 T, 0.3 T, 0.5 T, 0.7 T, respectively. (c) The energy barriers to be overcome when the skyrmions collapse to the ferromagnetic state. (d) Evolutions of spin textures of CoCl$_2$ monolayer with variational temperatures and magnetic fields and the right column is the enlarged views of typical skyrmion phase with zero temperature at 0T and 2.5T as outlined with yellow lines, respectively.
		}
	\end{center}
\end{figure}

Based on above discussions, to verify the reliability of above conclusions, we have found an applicable 2D material of CoCl$_2$ monolayer, as depicted in Fig. \ref{fig5}(a). the magnetic Co atoms with point group $C_{3v}$ form an HCP lattice are sandwiched by two nonmagnetic atomic planes consisting of Cl atoms. Additionally, the phonon dispersion is calculated to examine the stability of CoCl$_2$ in Fig. S6 (in the Supplemental Material\cite{SM}), and there is an exceedingly small, almost negligible imaginary frequency suggesting that it is dynamically stable. As presented in the band structures of Fig. S8 (in the Supplemental Material\cite{SM}), the CoCl$_2$ monolayer is a semiconductor with a sizeable bandgap of 2.32 eV.  Meanwhile, based on our DFT calculations, it is evident that the absence of DMI in the CoCl$_2$ monolayer can be attributed to its spatial inversion symmetry and we obtain the Hamiltonian parameters $J_1$ = 1.10 meV, $J_2$ = $-$0.38 meV and K = 0.04 meV. Subsequently, we investigate the skyrmions and phase transition through micromagnetic simulations. According to above discussions. The results of Fig. \ref{fig5}(b) show that, the relaxed skyrmions with $Q$ = 4, 3, 2, 1 can be obtained, suggesting that the frustrated coefficients $J_1$ , $J_2$ of CoCl$_2$ monolayer are expected to stabilize the skyrmions. Further then, the diameter of the skyrmions can be tunable by applying magnetic fields, and we find that the magnetic fields with 0.1 T, 0.3 T, 0.5 T,0.7 T can stabilize the skyrmion with the diameters of 8.53 nm, 6.11 nm, 3.64 nm, 1.00 nm, respectively. These small-size skyrmions with sub-10 nm diameters are technologically desirable due to their potential to significantly enhance the storage density of next-generation skyrmion-based information memory devices. While in the DMI systems, such small-size skyrmions often come from a large DMI\cite{RN5244,RN5241,RN3742} or the manipulation of magnetic field\cite{RN5231,RN1996,RN3047} to maintain their stability. Besides,  as shown in Fig. \ref{fig5}(c), the energy barriers are found between the skyrmion and FM state. The energy barriers for skyrmions with $Q$ = 1, 2, 3, 4 are 1.69 meV, 1.81 meV, 1.91 meV and 2.10 meV, respectively, which means that greater energy is required to annihilate a skyrmion with higher $Q$. Furthermore, the evolutions of spin textures of CoCl$_2$ with changing magnetic fields and temperatures are illustrated in Fig. \ref{fig5}(d), indicating the outline of skyrmions that can be generated in intrinsic CoCl$_2$ without external field. More interestingly, the isolated skyrmion occurs when the magnetic field surpasses 0.5 T at 0 K, then, as the magnetic field strength increases, more skyrmions are generated, eventually forming a skyrmion lattice (such as the spin textures at 2.5 T and 0 K). As the magnetic field slowly increases, which provides valuable guidance for the design of next-generation information memory devices.

\section{Conclusions}
In summary, based on the frustrated $J_1$-$J_2$ Heisenberg model and LLG equation, we systematically investigate the high-topological-number skyrmions and their phase transition in HCP lattice. Firstly, the spin textures of AFM, LD, skyrmion and FM are demonstrated within the given $J_1$-$J_2$ ranges. Then, the area of stabilized skyrmions with different $Q$ are summarized in the $J_1$-$J_2$ phase diagram. In addition, the helicity of skyrmion shows a linear dependence on the frustration strength, resulting in the tunable switch of skyrmion between N$\acute{e}$el and Bloch type. Similar to that of the DMI system, the diameter of skyrmion in frustrated HCP lattice can be effectively manipulated by external magnetic field. Finally, through the DFT calculations and micromagnetic simulations, we also predict the existence of high $Q$ skyrmions in frustrated CoCl$_2$ monolayer. Besides, the energy barriers of skyrmion collapse are calculated with the GNEB method, which indicates that the skyrmions with higher $Q$ possess a stronger topological protection. Our work provides guidance and reference for the investigations of the high-topological-number skyrmion and phase transition in frustrated system.

\begin{acknowledgments}
This work was supported by National Natural Science Foundation of China (Grant No.11804301), the Natural Science Foundation of Zhejiang Province (Grant No.LY21A040008), the Fundamental Research Funds of Zhejiang Sci-Tech University (Grant No.2021Q043-Y and LGYJY2021015).
\end{acknowledgments}
\bibliography{ref}

\begin{thebibliography}{65}%
\makeatletter
\providecommand \@ifxundefined [1]{%
 \@ifx{#1\undefined}
}%
\providecommand \@ifnum [1]{%
 \ifnum #1\expandafter \@firstoftwo
 \else \expandafter \@secondoftwo
 \fi
}%
\providecommand \@ifx [1]{%
 \ifx #1\expandafter \@firstoftwo
 \else \expandafter \@secondoftwo
 \fi
}%
\providecommand \natexlab [1]{#1}%
\providecommand \enquote  [1]{``#1''}%
\providecommand \bibnamefont  [1]{#1}%
\providecommand \bibfnamefont [1]{#1}%
\providecommand \citenamefont [1]{#1}%
\providecommand \href@noop [0]{\@secondoftwo}%
\providecommand \href [0]{\begingroup \@sanitize@url \@href}%
\providecommand \@href[1]{\@@startlink{#1}\@@href}%
\providecommand \@@href[1]{\endgroup#1\@@endlink}%
\providecommand \@sanitize@url [0]{\catcode `\\12\catcode `\$12\catcode
  `\&12\catcode `\#12\catcode `\^12\catcode `\_12\catcode `\%12\relax}%
\providecommand \@@startlink[1]{}%
\providecommand \@@endlink[0]{}%
\providecommand \url  [0]{\begingroup\@sanitize@url \@url }%
\providecommand \@url [1]{\endgroup\@href {#1}{\urlprefix }}%
\providecommand \urlprefix  [0]{URL }%
\providecommand \Eprint [0]{\href }%
\providecommand \doibase [0]{http://dx.doi.org/}%
\providecommand \selectlanguage [0]{\@gobble}%
\providecommand \bibinfo  [0]{\@secondoftwo}%
\providecommand \bibfield  [0]{\@secondoftwo}%
\providecommand \translation [1]{[#1]}%
\providecommand \BibitemOpen [0]{}%
\providecommand \bibitemStop [0]{}%
\providecommand \bibitemNoStop [0]{.\EOS\space}%
\providecommand \EOS [0]{\spacefactor3000\relax}%
\providecommand \BibitemShut  [1]{\csname bibitem#1\endcsname}%
\let\auto@bib@innerbib\@empty
\bibitem [{\citenamefont {Muhlbauer}\ \emph {et~al.}(2009)\citenamefont
  {Muhlbauer}, \citenamefont {Binz}, \citenamefont {Jonietz}, \citenamefont
  {Pfleiderer}, \citenamefont {Rosch}, \citenamefont {Neubauer}, \citenamefont
  {Georgii},\ and\ \citenamefont {Boni}}]{RN144}%
  \BibitemOpen
  \bibfield  {author} {\bibinfo {author} {\bibfnamefont {S.}~\bibnamefont
  {Muhlbauer}}, \bibinfo {author} {\bibfnamefont {B.}~\bibnamefont {Binz}},
  \bibinfo {author} {\bibfnamefont {F.}~\bibnamefont {Jonietz}}, \bibinfo
  {author} {\bibfnamefont {C.}~\bibnamefont {Pfleiderer}}, \bibinfo {author}
  {\bibfnamefont {A.}~\bibnamefont {Rosch}}, \bibinfo {author} {\bibfnamefont
  {A.}~\bibnamefont {Neubauer}}, \bibinfo {author} {\bibfnamefont
  {R.}~\bibnamefont {Georgii}}, \ and\ \bibinfo {author} {\bibfnamefont
  {P.}~\bibnamefont {Boni}},\ }\href {\doibase 10.1126/science.1166767}
  {\bibfield  {journal} {\bibinfo  {journal} {Science}\ }\textbf {\bibinfo
  {volume} {323}},\ \bibinfo {pages} {915} (\bibinfo {year}
  {2009})}\BibitemShut {NoStop}%
\bibitem [{\citenamefont {Yu}\ \emph {et~al.}(2010)\citenamefont {Yu},
  \citenamefont {Onose}, \citenamefont {Kanazawa}, \citenamefont {Park},
  \citenamefont {Han}, \citenamefont {Matsui}, \citenamefont {Nagaosa},\ and\
  \citenamefont {Tokura}}]{RN156}%
  \BibitemOpen
  \bibfield  {author} {\bibinfo {author} {\bibfnamefont {X.~Z.}\ \bibnamefont
  {Yu}}, \bibinfo {author} {\bibfnamefont {Y.}~\bibnamefont {Onose}}, \bibinfo
  {author} {\bibfnamefont {N.}~\bibnamefont {Kanazawa}}, \bibinfo {author}
  {\bibfnamefont {J.~H.}\ \bibnamefont {Park}}, \bibinfo {author}
  {\bibfnamefont {J.~H.}\ \bibnamefont {Han}}, \bibinfo {author} {\bibfnamefont
  {Y.}~\bibnamefont {Matsui}}, \bibinfo {author} {\bibfnamefont
  {N.}~\bibnamefont {Nagaosa}}, \ and\ \bibinfo {author} {\bibfnamefont
  {Y.}~\bibnamefont {Tokura}},\ }\href {\doibase 10.1038/nature09124}
  {\bibfield  {journal} {\bibinfo  {journal} {Nature}\ }\textbf {\bibinfo
  {volume} {465}},\ \bibinfo {pages} {901} (\bibinfo {year}
  {2010})}\BibitemShut {NoStop}%
\bibitem [{\citenamefont {Tokura}\ and\ \citenamefont {Kanazawa}(2021)}]{RN67}%
  \BibitemOpen
  \bibfield  {author} {\bibinfo {author} {\bibfnamefont {Y.}~\bibnamefont
  {Tokura}}\ and\ \bibinfo {author} {\bibfnamefont {N.}~\bibnamefont
  {Kanazawa}},\ }\href {\doibase 10.1021/acs.chemrev.0c00297} {\bibfield
  {journal} {\bibinfo  {journal} {Chem. Rev.}\ }\textbf {\bibinfo {volume}
  {121}},\ \bibinfo {pages} {2857} (\bibinfo {year} {2021})}\BibitemShut
  {NoStop}%
\bibitem [{\citenamefont {Yu~G}\ \emph {et~al.}(2017)\citenamefont {Yu~G},
  \citenamefont {Shao~Q}, \citenamefont {Yin~G}, \citenamefont {He~C},
  \citenamefont {Han~X},\ and\ \citenamefont {L}}]{RN5232}%
  \BibitemOpen
  \bibfield  {author} {\bibinfo {author} {\bibfnamefont {U.~P.}\ \bibnamefont
  {Yu~G}}, \bibinfo {author} {\bibfnamefont {W.~H.}\ \bibnamefont {Shao~Q}},
  \bibinfo {author} {\bibfnamefont {L.~X.}\ \bibnamefont {Yin~G}}, \bibinfo
  {author} {\bibfnamefont {J.~W.}\ \bibnamefont {He~C}}, \bibinfo {author}
  {\bibfnamefont {A.~P.~K.}\ \bibnamefont {Han~X}}, \ and\ \bibinfo {author}
  {\bibfnamefont {W.~K.}\ \bibnamefont {L}},\ }\href {\doibase
  10.1021/acs.nanolett.6b04010} {\bibfield  {journal} {\bibinfo  {journal}
  {Nano Lett.}\ }\textbf {\bibinfo {volume} {17}},\ \bibinfo {pages} {261}
  (\bibinfo {year} {2017})}\BibitemShut {NoStop}%
\bibitem [{\citenamefont {Kang}\ \emph {et~al.}(2017)\citenamefont {Kang},
  \citenamefont {Zheng}, \citenamefont {Huang}, \citenamefont {Zhang},
  \citenamefont {Lv}, \citenamefont {Zhou},\ and\ \citenamefont
  {Zhao}}]{RN5233}%
  \BibitemOpen
  \bibfield  {author} {\bibinfo {author} {\bibfnamefont {W.}~\bibnamefont
  {Kang}}, \bibinfo {author} {\bibfnamefont {C.}~\bibnamefont {Zheng}},
  \bibinfo {author} {\bibfnamefont {Y.}~\bibnamefont {Huang}}, \bibinfo
  {author} {\bibfnamefont {X.}~\bibnamefont {Zhang}}, \bibinfo {author}
  {\bibfnamefont {W.}~\bibnamefont {Lv}}, \bibinfo {author} {\bibfnamefont
  {Y.}~\bibnamefont {Zhou}}, \ and\ \bibinfo {author} {\bibfnamefont
  {W.}~\bibnamefont {Zhao}},\ }\href {\doibase 10.1109/TED.2017.2656140}
  {\bibfield  {journal} {\bibinfo  {journal} {IEEE Trans. Electron Devices}\
  }\textbf {\bibinfo {volume} {64}},\ \bibinfo {pages} {1060} (\bibinfo {year}
  {2017})}\BibitemShut {NoStop}%
\bibitem [{\citenamefont {Zhang}\ \emph {et~al.}(2015)\citenamefont {Zhang},
  \citenamefont {Zhou}, \citenamefont {Ezawa}, \citenamefont {Zhao},\ and\
  \citenamefont {Zhao}}]{RN5234}%
  \BibitemOpen
  \bibfield  {author} {\bibinfo {author} {\bibfnamefont {X.}~\bibnamefont
  {Zhang}}, \bibinfo {author} {\bibfnamefont {Y.}~\bibnamefont {Zhou}},
  \bibinfo {author} {\bibfnamefont {M.}~\bibnamefont {Ezawa}}, \bibinfo
  {author} {\bibfnamefont {G.~P.}\ \bibnamefont {Zhao}}, \ and\ \bibinfo
  {author} {\bibfnamefont {W.}~\bibnamefont {Zhao}},\ }\href {\doibase
  10.1038/srep11369} {\bibfield  {journal} {\bibinfo  {journal} {Sci. Rep.}\
  }\textbf {\bibinfo {volume} {5}},\ \bibinfo {pages} {11369} (\bibinfo {year}
  {2015})}\BibitemShut {NoStop}%
\bibitem [{\citenamefont {Kita}\ \emph {et~al.}(2012)\citenamefont {Kita},
  \citenamefont {Abraham}, \citenamefont {Gajek},\ and\ \citenamefont
  {Worledge}}]{RN5235}%
  \BibitemOpen
  \bibfield  {author} {\bibinfo {author} {\bibfnamefont {K.}~\bibnamefont
  {Kita}}, \bibinfo {author} {\bibfnamefont {D.~W.}\ \bibnamefont {Abraham}},
  \bibinfo {author} {\bibfnamefont {M.~J.}\ \bibnamefont {Gajek}}, \ and\
  \bibinfo {author} {\bibfnamefont {D.~C.}\ \bibnamefont {Worledge}},\ }\href
  {\doibase 10.1063/1.4745901} {\bibfield  {journal} {\bibinfo  {journal} {J.
  Appl. Phys.}\ }\textbf {\bibinfo {volume} {112}},\ \bibinfo {pages} {033919}
  (\bibinfo {year} {2012})}\BibitemShut {NoStop}%
\bibitem [{\citenamefont {Schellekens}\ \emph {et~al.}(2012)\citenamefont
  {Schellekens}, \citenamefont {van~den Brink}, \citenamefont {Franken},
  \citenamefont {Swagten},\ and\ \citenamefont {Koopmans}}]{RN5236}%
  \BibitemOpen
  \bibfield  {author} {\bibinfo {author} {\bibfnamefont {A.~J.}\ \bibnamefont
  {Schellekens}}, \bibinfo {author} {\bibfnamefont {A.}~\bibnamefont {van~den
  Brink}}, \bibinfo {author} {\bibfnamefont {J.~H.}\ \bibnamefont {Franken}},
  \bibinfo {author} {\bibfnamefont {H.~J.}\ \bibnamefont {Swagten}}, \ and\
  \bibinfo {author} {\bibfnamefont {B.}~\bibnamefont {Koopmans}},\ }\href
  {\doibase 10.1038/ncomms1848} {\bibfield  {journal} {\bibinfo  {journal}
  {Nat. Commun.}\ }\textbf {\bibinfo {volume} {3}},\ \bibinfo {pages} {847}
  (\bibinfo {year} {2012})}\BibitemShut {NoStop}%
\bibitem [{\citenamefont {Shiota}\ \emph {et~al.}(2011)\citenamefont {Shiota},
  \citenamefont {Murakami}, \citenamefont {Bonell}, \citenamefont {Nozaki},
  \citenamefont {Shinjo},\ and\ \citenamefont {Suzuki}}]{RN5237}%
  \BibitemOpen
  \bibfield  {author} {\bibinfo {author} {\bibfnamefont {Y.}~\bibnamefont
  {Shiota}}, \bibinfo {author} {\bibfnamefont {S.}~\bibnamefont {Murakami}},
  \bibinfo {author} {\bibfnamefont {F.}~\bibnamefont {Bonell}}, \bibinfo
  {author} {\bibfnamefont {T.}~\bibnamefont {Nozaki}}, \bibinfo {author}
  {\bibfnamefont {T.}~\bibnamefont {Shinjo}}, \ and\ \bibinfo {author}
  {\bibfnamefont {Y.}~\bibnamefont {Suzuki}},\ }\href {\doibase
  10.1143/APEX.4.043005} {\bibfield  {journal} {\bibinfo  {journal} {Appl.
  Phys. Express}\ }\textbf {\bibinfo {volume} {4}},\ \bibinfo {pages} {043005}
  (\bibinfo {year} {2011})}\BibitemShut {NoStop}%
\bibitem [{\citenamefont {Huang}\ \emph {et~al.}(2017)\citenamefont {Huang},
  \citenamefont {Kang}, \citenamefont {Zhang}, \citenamefont {Zhou},\ and\
  \citenamefont {Zhao}}]{RN5238}%
  \BibitemOpen
  \bibfield  {author} {\bibinfo {author} {\bibfnamefont {Y.}~\bibnamefont
  {Huang}}, \bibinfo {author} {\bibfnamefont {W.}~\bibnamefont {Kang}},
  \bibinfo {author} {\bibfnamefont {X.}~\bibnamefont {Zhang}}, \bibinfo
  {author} {\bibfnamefont {Y.}~\bibnamefont {Zhou}}, \ and\ \bibinfo {author}
  {\bibfnamefont {W.}~\bibnamefont {Zhao}},\ }\href {\doibase
  10.1088/1361-6528/aa5838} {\bibfield  {journal} {\bibinfo  {journal}
  {Nanotechnology}\ }\textbf {\bibinfo {volume} {28}},\ \bibinfo {pages}
  {08LT02} (\bibinfo {year} {2017})}\BibitemShut {NoStop}%
\bibitem [{\citenamefont {Li}\ \emph {et~al.}(2017)\citenamefont {Li},
  \citenamefont {Kang}, \citenamefont {Huang}, \citenamefont {Zhang},
  \citenamefont {Zhou},\ and\ \citenamefont {Zhao}}]{RN5239}%
  \BibitemOpen
  \bibfield  {author} {\bibinfo {author} {\bibfnamefont {S.}~\bibnamefont
  {Li}}, \bibinfo {author} {\bibfnamefont {W.}~\bibnamefont {Kang}}, \bibinfo
  {author} {\bibfnamefont {Y.}~\bibnamefont {Huang}}, \bibinfo {author}
  {\bibfnamefont {X.}~\bibnamefont {Zhang}}, \bibinfo {author} {\bibfnamefont
  {Y.}~\bibnamefont {Zhou}}, \ and\ \bibinfo {author} {\bibfnamefont
  {W.}~\bibnamefont {Zhao}},\ }\href {\doibase 10.1088/1361-6528/aa7af5}
  {\bibfield  {journal} {\bibinfo  {journal} {Nanotechnology}\ }\textbf
  {\bibinfo {volume} {28}},\ \bibinfo {pages} {31LT01} (\bibinfo {year}
  {2017})}\BibitemShut {NoStop}%
\bibitem [{\citenamefont {Yao}\ \emph {et~al.}(2023)\citenamefont {Yao},
  \citenamefont {Hu},\ and\ \citenamefont {Dong}}]{RN5293}%
  \BibitemOpen
  \bibfield  {author} {\bibinfo {author} {\bibfnamefont {X.}~\bibnamefont
  {Yao}}, \bibinfo {author} {\bibfnamefont {D.}~\bibnamefont {Hu}}, \ and\
  \bibinfo {author} {\bibfnamefont {S.}~\bibnamefont {Dong}},\ }\href {\doibase
  https://doi.org/10.1016/j.isci.2023.106311} {\bibfield  {journal} {\bibinfo
  {journal} {iScience}\ }\textbf {\bibinfo {volume} {26}},\ \bibinfo {pages}
  {106311} (\bibinfo {year} {2023})}\BibitemShut {NoStop}%
\bibitem [{\citenamefont {Dzyaloshinsky}(1958)}]{RN130}%
  \BibitemOpen
  \bibfield  {author} {\bibinfo {author} {\bibfnamefont {I.}~\bibnamefont
  {Dzyaloshinsky}},\ }\href {\doibase
  https://doi.org/10.1016/0022-3697(58)90076-3} {\bibfield  {journal} {\bibinfo
   {journal} {J. Phys. Chem. Solids}\ }\textbf {\bibinfo {volume} {4}},\
  \bibinfo {pages} {241} (\bibinfo {year} {1958})}\BibitemShut {NoStop}%
\bibitem [{\citenamefont {Moriya}(1960)}]{RN131}%
  \BibitemOpen
  \bibfield  {author} {\bibinfo {author} {\bibfnamefont {T.}~\bibnamefont
  {Moriya}},\ }\href {\doibase 10.1103/PhysRev.120.91} {\bibfield  {journal}
  {\bibinfo  {journal} {Phys. Rev.}\ }\textbf {\bibinfo {volume} {120}},\
  \bibinfo {pages} {91} (\bibinfo {year} {1960})}\BibitemShut {NoStop}%
\bibitem [{\citenamefont {Ga}\ \emph {et~al.}(2022)\citenamefont {Ga},
  \citenamefont {Cui}, \citenamefont {Zhu}, \citenamefont {Yu}, \citenamefont
  {Wang}, \citenamefont {Liang},\ and\ \citenamefont {Yang}}]{RN3742}%
  \BibitemOpen
  \bibfield  {author} {\bibinfo {author} {\bibfnamefont {Y.}~\bibnamefont
  {Ga}}, \bibinfo {author} {\bibfnamefont {Q.}~\bibnamefont {Cui}}, \bibinfo
  {author} {\bibfnamefont {Y.}~\bibnamefont {Zhu}}, \bibinfo {author}
  {\bibfnamefont {D.}~\bibnamefont {Yu}}, \bibinfo {author} {\bibfnamefont
  {L.}~\bibnamefont {Wang}}, \bibinfo {author} {\bibfnamefont {J.}~\bibnamefont
  {Liang}}, \ and\ \bibinfo {author} {\bibfnamefont {H.}~\bibnamefont {Yang}},\
  }\href {\doibase 10.1038/s41524-022-00809-4} {\bibfield  {journal} {\bibinfo
  {journal} {npj Comput. Mater.}\ }\textbf {\bibinfo {volume} {8}},\ \bibinfo
  {pages} {128} (\bibinfo {year} {2022})}\BibitemShut {NoStop}%
\bibitem [{\citenamefont {Cui}\ \emph {et~al.}(2020)\citenamefont {Cui},
  \citenamefont {Liang}, \citenamefont {Shao}, \citenamefont {Cui},\ and\
  \citenamefont {Yang}}]{RN31}%
  \BibitemOpen
  \bibfield  {author} {\bibinfo {author} {\bibfnamefont {Q.}~\bibnamefont
  {Cui}}, \bibinfo {author} {\bibfnamefont {J.}~\bibnamefont {Liang}}, \bibinfo
  {author} {\bibfnamefont {Z.}~\bibnamefont {Shao}}, \bibinfo {author}
  {\bibfnamefont {P.}~\bibnamefont {Cui}}, \ and\ \bibinfo {author}
  {\bibfnamefont {H.}~\bibnamefont {Yang}},\ }\href {\doibase
  10.1103/PhysRevB.102.094425} {\bibfield  {journal} {\bibinfo  {journal}
  {Phys. Rev. B}\ }\textbf {\bibinfo {volume} {102}},\ \bibinfo {pages}
  {094425} (\bibinfo {year} {2020})}\BibitemShut {NoStop}%
\bibitem [{\citenamefont {Nagaosa}(2013)}]{RN3688}%
  \BibitemOpen
  \bibfield  {author} {\bibinfo {author} {\bibfnamefont {Y.}~\bibnamefont
  {Nagaosa}, \bibfnamefont {N.~Tokura}},\ }\href {\doibase
  10.1038/nnano.2013.243} {\bibfield  {journal} {\bibinfo  {journal} {Nat.
  Nanotechnol.}\ }\textbf {\bibinfo {volume} {8}},\ \bibinfo {pages} {899}
  (\bibinfo {year} {2013})}\BibitemShut {NoStop}%
\bibitem [{\citenamefont {Xu}\ \emph {et~al.}(2020)\citenamefont {Xu},
  \citenamefont {Feng}, \citenamefont {Prokhorenko}, \citenamefont {Nahas},
  \citenamefont {Xiang},\ and\ \citenamefont {Bellaiche}}]{RN1041}%
  \BibitemOpen
  \bibfield  {author} {\bibinfo {author} {\bibfnamefont {C.}~\bibnamefont
  {Xu}}, \bibinfo {author} {\bibfnamefont {J.}~\bibnamefont {Feng}}, \bibinfo
  {author} {\bibfnamefont {S.}~\bibnamefont {Prokhorenko}}, \bibinfo {author}
  {\bibfnamefont {Y.}~\bibnamefont {Nahas}}, \bibinfo {author} {\bibfnamefont
  {H.}~\bibnamefont {Xiang}}, \ and\ \bibinfo {author} {\bibfnamefont
  {L.}~\bibnamefont {Bellaiche}},\ }\href {\doibase
  10.1103/PhysRevB.101.060404} {\bibfield  {journal} {\bibinfo  {journal}
  {Phys. Rev. B}\ }\textbf {\bibinfo {volume} {101}},\ \bibinfo {pages}
  {060404(R)} (\bibinfo {year} {2020})}\BibitemShut {NoStop}%
\bibitem [{\citenamefont {Zhang}\ \emph {et~al.}(2016)\citenamefont {Zhang},
  \citenamefont {Zhou},\ and\ \citenamefont {Ezawa}}]{RN199}%
  \BibitemOpen
  \bibfield  {author} {\bibinfo {author} {\bibfnamefont {X.}~\bibnamefont
  {Zhang}}, \bibinfo {author} {\bibfnamefont {Y.}~\bibnamefont {Zhou}}, \ and\
  \bibinfo {author} {\bibfnamefont {M.}~\bibnamefont {Ezawa}},\ }\href
  {\doibase 10.1103/PhysRevB.93.024415} {\bibfield  {journal} {\bibinfo
  {journal} {Phys. Rev. B}\ }\textbf {\bibinfo {volume} {93}},\ \bibinfo
  {pages} {024415} (\bibinfo {year} {2016})}\BibitemShut {NoStop}%
\bibitem [{\citenamefont {Zhang}\ \emph {et~al.}(2017)\citenamefont {Zhang},
  \citenamefont {Xia}, \citenamefont {Zhou}, \citenamefont {Liu}, \citenamefont
  {Zhang},\ and\ \citenamefont {Ezawa}}]{RN3326}%
  \BibitemOpen
  \bibfield  {author} {\bibinfo {author} {\bibfnamefont {X.}~\bibnamefont
  {Zhang}}, \bibinfo {author} {\bibfnamefont {J.}~\bibnamefont {Xia}}, \bibinfo
  {author} {\bibfnamefont {Y.}~\bibnamefont {Zhou}}, \bibinfo {author}
  {\bibfnamefont {X.}~\bibnamefont {Liu}}, \bibinfo {author} {\bibfnamefont
  {H.}~\bibnamefont {Zhang}}, \ and\ \bibinfo {author} {\bibfnamefont
  {M.}~\bibnamefont {Ezawa}},\ }\href {\doibase 10.1038/s41467-017-01785-w}
  {\bibfield  {journal} {\bibinfo  {journal} {Nat. Commun.}\ }\textbf {\bibinfo
  {volume} {8}},\ \bibinfo {pages} {1717} (\bibinfo {year} {2017})}\BibitemShut
  {NoStop}%
\bibitem [{\citenamefont {Leonov}\ and\ \citenamefont
  {Mostovoy}(2015)}]{RN3020}%
  \BibitemOpen
  \bibfield  {author} {\bibinfo {author} {\bibfnamefont {A.~O.}\ \bibnamefont
  {Leonov}}\ and\ \bibinfo {author} {\bibfnamefont {M.}~\bibnamefont
  {Mostovoy}},\ }\href {\doibase 10.1038/ncomms9275} {\bibfield  {journal}
  {\bibinfo  {journal} {Nat. Commun.}\ }\textbf {\bibinfo {volume} {6}},\
  \bibinfo {pages} {8275} (\bibinfo {year} {2015})}\BibitemShut {NoStop}%
\bibitem [{\citenamefont {Liang}\ \emph {et~al.}(2018)\citenamefont {Liang},
  \citenamefont {Yu}, \citenamefont {Chen}, \citenamefont {Qin}, \citenamefont
  {Zeng}, \citenamefont {Lu}, \citenamefont {Gao},\ and\ \citenamefont
  {Liu}}]{RN3728}%
  \BibitemOpen
  \bibfield  {author} {\bibinfo {author} {\bibfnamefont {J.~J.}\ \bibnamefont
  {Liang}}, \bibinfo {author} {\bibfnamefont {J.~H.}\ \bibnamefont {Yu}},
  \bibinfo {author} {\bibfnamefont {J.}~\bibnamefont {Chen}}, \bibinfo {author}
  {\bibfnamefont {M.~H.}\ \bibnamefont {Qin}}, \bibinfo {author} {\bibfnamefont
  {M.}~\bibnamefont {Zeng}}, \bibinfo {author} {\bibfnamefont {X.~B.}\
  \bibnamefont {Lu}}, \bibinfo {author} {\bibfnamefont {X.~S.}\ \bibnamefont
  {Gao}}, \ and\ \bibinfo {author} {\bibfnamefont {J.}~\bibnamefont {Liu}},\
  }\href {\doibase 10.1088/1367-2630/aac24c} {\bibfield  {journal} {\bibinfo
  {journal} {New J. Phys.}\ }\textbf {\bibinfo {volume} {20}},\ \bibinfo
  {pages} {053037} (\bibinfo {year} {2018})}\BibitemShut {NoStop}%
\bibitem [{\citenamefont {Hou}\ \emph {et~al.}(2020)\citenamefont {Hou},
  \citenamefont {Zhang}, \citenamefont {Zhang}, \citenamefont {Xu},
  \citenamefont {Xia}, \citenamefont {Ding}, \citenamefont {Li}, \citenamefont
  {Zhang}, \citenamefont {Batra}, \citenamefont {Costa}, \citenamefont {Liu},
  \citenamefont {Wu}, \citenamefont {Ezawa}, \citenamefont {Liu}, \citenamefont
  {Zhou}, \citenamefont {Zhang},\ and\ \citenamefont {Wang}}]{RN3103}%
  \BibitemOpen
  \bibfield  {author} {\bibinfo {author} {\bibfnamefont {Z.}~\bibnamefont
  {Hou}}, \bibinfo {author} {\bibfnamefont {Q.}~\bibnamefont {Zhang}}, \bibinfo
  {author} {\bibfnamefont {X.}~\bibnamefont {Zhang}}, \bibinfo {author}
  {\bibfnamefont {G.}~\bibnamefont {Xu}}, \bibinfo {author} {\bibfnamefont
  {J.}~\bibnamefont {Xia}}, \bibinfo {author} {\bibfnamefont {B.}~\bibnamefont
  {Ding}}, \bibinfo {author} {\bibfnamefont {H.}~\bibnamefont {Li}}, \bibinfo
  {author} {\bibfnamefont {S.}~\bibnamefont {Zhang}}, \bibinfo {author}
  {\bibfnamefont {N.~M.}\ \bibnamefont {Batra}}, \bibinfo {author}
  {\bibfnamefont {P.}~\bibnamefont {Costa}}, \bibinfo {author} {\bibfnamefont
  {E.}~\bibnamefont {Liu}}, \bibinfo {author} {\bibfnamefont {G.}~\bibnamefont
  {Wu}}, \bibinfo {author} {\bibfnamefont {M.}~\bibnamefont {Ezawa}}, \bibinfo
  {author} {\bibfnamefont {X.}~\bibnamefont {Liu}}, \bibinfo {author}
  {\bibfnamefont {Y.}~\bibnamefont {Zhou}}, \bibinfo {author} {\bibfnamefont
  {X.}~\bibnamefont {Zhang}}, \ and\ \bibinfo {author} {\bibfnamefont
  {W.}~\bibnamefont {Wang}},\ }\href {\doibase 10.1002/adma.201904815}
  {\bibfield  {journal} {\bibinfo  {journal} {Adv. Mater.}\ }\textbf {\bibinfo
  {volume} {32}},\ \bibinfo {pages} {e1904815} (\bibinfo {year}
  {2020})}\BibitemShut {NoStop}%
\bibitem [{\citenamefont {Miyata}\ \emph {et~al.}(2022)\citenamefont {Miyata},
  \citenamefont {Ohe},\ and\ \citenamefont {Tatara}}]{RN152}%
  \BibitemOpen
  \bibfield  {author} {\bibinfo {author} {\bibfnamefont {M.}~\bibnamefont
  {Miyata}}, \bibinfo {author} {\bibfnamefont {J.~I.}\ \bibnamefont {Ohe}}, \
  and\ \bibinfo {author} {\bibfnamefont {G.}~\bibnamefont {Tatara}},\ }\href
  {\doibase 10.1103/PhysRevApplied.18.014075} {\bibfield  {journal} {\bibinfo
  {journal} {Phys. Rev. Appl.}\ }\textbf {\bibinfo {volume} {18}},\ \bibinfo
  {pages} {014075} (\bibinfo {year} {2022})}\BibitemShut {NoStop}%
\bibitem [{\citenamefont {Leonov}\ and\ \citenamefont
  {Mostovoy}(2017)}]{RN116}%
  \BibitemOpen
  \bibfield  {author} {\bibinfo {author} {\bibfnamefont {A.~O.}\ \bibnamefont
  {Leonov}}\ and\ \bibinfo {author} {\bibfnamefont {M.}~\bibnamefont
  {Mostovoy}},\ }\href {\doibase 10.1038/ncomms14394} {\bibfield  {journal}
  {\bibinfo  {journal} {Nat. Commun.}\ }\textbf {\bibinfo {volume} {8}},\
  \bibinfo {pages} {14394} (\bibinfo {year} {2017})}\BibitemShut {NoStop}%
\bibitem [{\citenamefont {Yao}\ and\ \citenamefont {Dong}(2022)}]{RN5251}%
  \BibitemOpen
  \bibfield  {author} {\bibinfo {author} {\bibfnamefont {X.}~\bibnamefont
  {Yao}}\ and\ \bibinfo {author} {\bibfnamefont {S.}~\bibnamefont {Dong}},\
  }\href {\doibase 10.1103/PhysRevB.105.014444} {\bibfield  {journal} {\bibinfo
   {journal} {Phys. Rev. B}\ }\textbf {\bibinfo {volume} {105}},\ \bibinfo
  {pages} {014444} (\bibinfo {year} {2022})}\BibitemShut {NoStop}%
\bibitem [{\citenamefont {Zhang}\ \emph {et~al.}(2020)\citenamefont {Zhang},
  \citenamefont {Xia}, \citenamefont {Shen}, \citenamefont {Ezawa},
  \citenamefont {Tretiakov}, \citenamefont {Zhao}, \citenamefont {Liu},\ and\
  \citenamefont {Zhou}}]{RN151}%
  \BibitemOpen
  \bibfield  {author} {\bibinfo {author} {\bibfnamefont {X.}~\bibnamefont
  {Zhang}}, \bibinfo {author} {\bibfnamefont {J.}~\bibnamefont {Xia}}, \bibinfo
  {author} {\bibfnamefont {L.}~\bibnamefont {Shen}}, \bibinfo {author}
  {\bibfnamefont {M.}~\bibnamefont {Ezawa}}, \bibinfo {author} {\bibfnamefont
  {O.~A.}\ \bibnamefont {Tretiakov}}, \bibinfo {author} {\bibfnamefont
  {G.}~\bibnamefont {Zhao}}, \bibinfo {author} {\bibfnamefont {X.}~\bibnamefont
  {Liu}}, \ and\ \bibinfo {author} {\bibfnamefont {Y.}~\bibnamefont {Zhou}},\
  }\href {\doibase 10.1103/PhysRevB.101.144435} {\bibfield  {journal} {\bibinfo
   {journal} {Phys. Rev. B}\ }\textbf {\bibinfo {volume} {101}},\ \bibinfo
  {pages} {144435} (\bibinfo {year} {2020})}\BibitemShut {NoStop}%
\bibitem [{\citenamefont {Yao}\ \emph {et~al.}(2020)\citenamefont {Yao},
  \citenamefont {Chen},\ and\ \citenamefont {Dong}}]{RN3741}%
  \BibitemOpen
  \bibfield  {author} {\bibinfo {author} {\bibfnamefont {X.}~\bibnamefont
  {Yao}}, \bibinfo {author} {\bibfnamefont {J.}~\bibnamefont {Chen}}, \ and\
  \bibinfo {author} {\bibfnamefont {S.}~\bibnamefont {Dong}},\ }\href {\doibase
  10.1088/1367-2630/aba1b3} {\bibfield  {journal} {\bibinfo  {journal} {New J.
  Phys.}\ }\textbf {\bibinfo {volume} {22}},\ \bibinfo {pages} {083032}
  (\bibinfo {year} {2020})}\BibitemShut {NoStop}%
\bibitem [{\citenamefont {Yuan}\ \emph
  {et~al.}(2017{\natexlab{a}})\citenamefont {Yuan}, \citenamefont {Gomonay},\
  and\ \citenamefont {Klaui}}]{RN153}%
  \BibitemOpen
  \bibfield  {author} {\bibinfo {author} {\bibfnamefont {H.~Y.}\ \bibnamefont
  {Yuan}}, \bibinfo {author} {\bibfnamefont {O.}~\bibnamefont {Gomonay}}, \
  and\ \bibinfo {author} {\bibfnamefont {M.}~\bibnamefont {Klaui}},\ }\href
  {\doibase 10.1103/PhysRevB.96.134415} {\bibfield  {journal} {\bibinfo
  {journal} {Phys. Rev. B}\ }\textbf {\bibinfo {volume} {96}},\ \bibinfo
  {pages} {134415} (\bibinfo {year} {2017}{\natexlab{a}})}\BibitemShut
  {NoStop}%
\bibitem [{\citenamefont {Xia}\ \emph {et~al.}(2023)\citenamefont {Xia},
  \citenamefont {Zhang}, \citenamefont {Liu}, \citenamefont {Zhou},\ and\
  \citenamefont {Ezawa}}]{RN105}%
  \BibitemOpen
  \bibfield  {author} {\bibinfo {author} {\bibfnamefont {J.}~\bibnamefont
  {Xia}}, \bibinfo {author} {\bibfnamefont {X.}~\bibnamefont {Zhang}}, \bibinfo
  {author} {\bibfnamefont {X.}~\bibnamefont {Liu}}, \bibinfo {author}
  {\bibfnamefont {Y.}~\bibnamefont {Zhou}}, \ and\ \bibinfo {author}
  {\bibfnamefont {M.}~\bibnamefont {Ezawa}},\ }\href {\doibase
  10.1103/PhysRevLett.130.106701} {\bibfield  {journal} {\bibinfo  {journal}
  {Phys. Rev. Lett.}\ }\textbf {\bibinfo {volume} {130}},\ \bibinfo {pages}
  {106701} (\bibinfo {year} {2023})}\BibitemShut {NoStop}%
\bibitem [{\citenamefont {Berge}\ \emph {et~al.}(1986)\citenamefont {Berge},
  \citenamefont {Diep}, \citenamefont {Ghazali},\ and\ \citenamefont
  {Lallemand}}]{RN81}%
  \BibitemOpen
  \bibfield  {author} {\bibinfo {author} {\bibfnamefont {B.}~\bibnamefont
  {Berge}}, \bibinfo {author} {\bibfnamefont {H.~T.}\ \bibnamefont {Diep}},
  \bibinfo {author} {\bibfnamefont {A.}~\bibnamefont {Ghazali}}, \ and\
  \bibinfo {author} {\bibfnamefont {P.}~\bibnamefont {Lallemand}},\ }\href
  {\doibase 10.1103/PhysRevB.34.3177} {\bibfield  {journal} {\bibinfo
  {journal} {Phys. Rev. B}\ }\textbf {\bibinfo {volume} {34}},\ \bibinfo
  {pages} {3177} (\bibinfo {year} {1986})}\BibitemShut {NoStop}%
\bibitem [{\citenamefont {Schlickum}\ \emph {et~al.}(2004)\citenamefont
  {Schlickum}, \citenamefont {Janke-Gilman}, \citenamefont {Wulfhekel},\ and\
  \citenamefont {Kirschner}}]{RN80}%
  \BibitemOpen
  \bibfield  {author} {\bibinfo {author} {\bibfnamefont {U.}~\bibnamefont
  {Schlickum}}, \bibinfo {author} {\bibfnamefont {N.}~\bibnamefont
  {Janke-Gilman}}, \bibinfo {author} {\bibfnamefont {W.}~\bibnamefont
  {Wulfhekel}}, \ and\ \bibinfo {author} {\bibfnamefont {J.}~\bibnamefont
  {Kirschner}},\ }\href {\doibase 10.1103/PhysRevLett.92.107203} {\bibfield
  {journal} {\bibinfo  {journal} {Phys. Rev. Lett.}\ }\textbf {\bibinfo
  {volume} {92}},\ \bibinfo {pages} {107203} (\bibinfo {year}
  {2004})}\BibitemShut {NoStop}%
\bibitem [{\citenamefont {Palmer}\ and\ \citenamefont {Chalker}(2000)}]{RN82}%
  \BibitemOpen
  \bibfield  {author} {\bibinfo {author} {\bibfnamefont {S.~E.}\ \bibnamefont
  {Palmer}}\ and\ \bibinfo {author} {\bibfnamefont {J.~T.}\ \bibnamefont
  {Chalker}},\ }\href {\doibase 10.1103/PhysRevB.62.488} {\bibfield  {journal}
  {\bibinfo  {journal} {Phys. Rev. B}\ }\textbf {\bibinfo {volume} {62}},\
  \bibinfo {pages} {488} (\bibinfo {year} {2000})}\BibitemShut {NoStop}%
\bibitem [{\citenamefont {Yuan}\ \emph
  {et~al.}(2017{\natexlab{b}})\citenamefont {Yuan}, \citenamefont {Gomonay},\
  and\ \citenamefont {Klaui}}]{RN79}%
  \BibitemOpen
  \bibfield  {author} {\bibinfo {author} {\bibfnamefont {H.~Y.}\ \bibnamefont
  {Yuan}}, \bibinfo {author} {\bibfnamefont {O.}~\bibnamefont {Gomonay}}, \
  and\ \bibinfo {author} {\bibfnamefont {M.}~\bibnamefont {Klaui}},\ }\href
  {\doibase 10.1103/PhysRevB.96.134415} {\bibfield  {journal} {\bibinfo
  {journal} {Phys. Rev. B}\ }\textbf {\bibinfo {volume} {96}},\ \bibinfo
  {pages} {134415} (\bibinfo {year} {2017}{\natexlab{b}})}\BibitemShut
  {NoStop}%
\bibitem [{\citenamefont {Sharafullin}\ and\ \citenamefont
  {Diep}(2019)}]{RN135}%
  \BibitemOpen
  \bibfield  {author} {\bibinfo {author} {\bibfnamefont {I.~F.}\ \bibnamefont
  {Sharafullin}}\ and\ \bibinfo {author} {\bibfnamefont {H.~T.}\ \bibnamefont
  {Diep}},\ }\href {\doibase 10.3390/sym12010026} {\bibfield  {journal}
  {\bibinfo  {journal} {Symmetry}\ }\textbf {\bibinfo {volume} {12}},\ \bibinfo
  {pages} {26} (\bibinfo {year} {2019})}\BibitemShut {NoStop}%
\bibitem [{\citenamefont {Kleiber}\ \emph {et~al.}(2000)\citenamefont
  {Kleiber}, \citenamefont {Bode}, \citenamefont {Ravlic},\ and\ \citenamefont
  {Wiesendanger}}]{RN84}%
  \BibitemOpen
  \bibfield  {author} {\bibinfo {author} {\bibfnamefont {M.}~\bibnamefont
  {Kleiber}}, \bibinfo {author} {\bibfnamefont {M.}~\bibnamefont {Bode}},
  \bibinfo {author} {\bibfnamefont {R.}~\bibnamefont {Ravlic}}, \ and\ \bibinfo
  {author} {\bibfnamefont {R.}~\bibnamefont {Wiesendanger}},\ }\href {\doibase
  10.1103/PhysRevLett.85.4606} {\bibfield  {journal} {\bibinfo  {journal}
  {Phys. Rev. Lett.}\ }\textbf {\bibinfo {volume} {85}},\ \bibinfo {pages}
  {4606} (\bibinfo {year} {2000})}\BibitemShut {NoStop}%
\bibitem [{\citenamefont {Owerre}(2017)}]{RN85}%
  \BibitemOpen
  \bibfield  {author} {\bibinfo {author} {\bibfnamefont {S.~A.}\ \bibnamefont
  {Owerre}},\ }\href {\doibase 10.1103/PhysRevB.95.014422} {\bibfield
  {journal} {\bibinfo  {journal} {Phys. Rev. B}\ }\textbf {\bibinfo {volume}
  {95}},\ \bibinfo {pages} {014422} (\bibinfo {year} {2017})}\BibitemShut
  {NoStop}%
\bibitem [{\citenamefont {Shea}\ \emph {et~al.}(1999)\citenamefont {Shea},
  \citenamefont {Onuchic},\ and\ \citenamefont {Brooks}}]{RN83}%
  \BibitemOpen
  \bibfield  {author} {\bibinfo {author} {\bibfnamefont {J.~E.}\ \bibnamefont
  {Shea}}, \bibinfo {author} {\bibfnamefont {J.~N.}\ \bibnamefont {Onuchic}}, \
  and\ \bibinfo {author} {\bibfnamefont {r.}~\bibnamefont {Brooks},
  \bibfnamefont {C.~L.}},\ }\href {\doibase 10.1073/pnas.96.22.12512}
  {\bibfield  {journal} {\bibinfo  {journal} {Proc. Natl. Acad. Sci. U.S.A.}\
  }\textbf {\bibinfo {volume} {96}},\ \bibinfo {pages} {12512} (\bibinfo {year}
  {1999})}\BibitemShut {NoStop}%
\bibitem [{\citenamefont {Westerhout}\ \emph {et~al.}(2020)\citenamefont
  {Westerhout}, \citenamefont {Astrakhantsev}, \citenamefont {Tikhonov},
  \citenamefont {Katsnelson},\ and\ \citenamefont {Bagrov}}]{RN155}%
  \BibitemOpen
  \bibfield  {author} {\bibinfo {author} {\bibfnamefont {T.}~\bibnamefont
  {Westerhout}}, \bibinfo {author} {\bibfnamefont {N.}~\bibnamefont
  {Astrakhantsev}}, \bibinfo {author} {\bibfnamefont {K.~S.}\ \bibnamefont
  {Tikhonov}}, \bibinfo {author} {\bibfnamefont {M.~I.}\ \bibnamefont
  {Katsnelson}}, \ and\ \bibinfo {author} {\bibfnamefont {A.~A.}\ \bibnamefont
  {Bagrov}},\ }\href {\doibase ARTN 1593 10.1038/s41467-020-15402-w} {\bibfield
   {journal} {\bibinfo  {journal} {Nat. Commun.}\ }\textbf {\bibinfo {volume}
  {11}},\ \bibinfo {pages} {1593} (\bibinfo {year} {2020})}\BibitemShut
  {NoStop}%
\bibitem [{\citenamefont {Mosadeq}\ \emph {et~al.}(2011)\citenamefont
  {Mosadeq}, \citenamefont {Shahbazi},\ and\ \citenamefont {Jafari}}]{RN154}%
  \BibitemOpen
  \bibfield  {author} {\bibinfo {author} {\bibfnamefont {H.}~\bibnamefont
  {Mosadeq}}, \bibinfo {author} {\bibfnamefont {F.}~\bibnamefont {Shahbazi}}, \
  and\ \bibinfo {author} {\bibfnamefont {S.~A.}\ \bibnamefont {Jafari}},\
  }\href {\doibase 10.1088/0953-8984/23/22/226006} {\bibfield  {journal}
  {\bibinfo  {journal} {J. Phys. Condens. Matter}\ }\textbf {\bibinfo {volume}
  {23}},\ \bibinfo {pages} {226006} (\bibinfo {year} {2011})}\BibitemShut
  {NoStop}%
\bibitem [{\citenamefont {Muller}\ \emph {et~al.}(2019)\citenamefont {Muller},
  \citenamefont {Hoffmann}, \citenamefont {Disselkamp}, \citenamefont
  {Schurhoff}, \citenamefont {Mavros}, \citenamefont {Sallermann},
  \citenamefont {Kiselev}, \citenamefont {Jonsson},\ and\ \citenamefont
  {Blugel}}]{RN1055}%
  \BibitemOpen
  \bibfield  {author} {\bibinfo {author} {\bibfnamefont {G.~P.}\ \bibnamefont
  {Muller}}, \bibinfo {author} {\bibfnamefont {M.}~\bibnamefont {Hoffmann}},
  \bibinfo {author} {\bibfnamefont {C.}~\bibnamefont {Disselkamp}}, \bibinfo
  {author} {\bibfnamefont {D.}~\bibnamefont {Schurhoff}}, \bibinfo {author}
  {\bibfnamefont {S.}~\bibnamefont {Mavros}}, \bibinfo {author} {\bibfnamefont
  {M.}~\bibnamefont {Sallermann}}, \bibinfo {author} {\bibfnamefont {N.~S.}\
  \bibnamefont {Kiselev}}, \bibinfo {author} {\bibfnamefont {H.}~\bibnamefont
  {Jonsson}}, \ and\ \bibinfo {author} {\bibfnamefont {S.}~\bibnamefont
  {Blugel}},\ }\href {\doibase 10.1103/PhysRevB.99.224414} {\bibfield
  {journal} {\bibinfo  {journal} {Phys. Rev. B}\ }\textbf {\bibinfo {volume}
  {99}},\ \bibinfo {pages} {224414} (\bibinfo {year} {2019})}\BibitemShut
  {NoStop}%
\bibitem [{\citenamefont {Landau}\ and\ \citenamefont
  {Lifshitz}(1935)}]{RN1984}%
  \BibitemOpen
  \bibfield  {author} {\bibinfo {author} {\bibfnamefont {L.~D.}\ \bibnamefont
  {Landau}}\ and\ \bibinfo {author} {\bibfnamefont {E.~M.}\ \bibnamefont
  {Lifshitz}},\ }\href {https://api.semanticscholar.org/CorpusID:26634163}
  {\bibfield  {journal} {\bibinfo  {journal} {Phys. Z. Sowjetunion}\ }\textbf
  {\bibinfo {volume} {8}},\ \bibinfo {pages} {51} (\bibinfo {year}
  {1935})}\BibitemShut {NoStop}%
\bibitem [{\citenamefont {Gilbert}(2004)}]{RN1985}%
  \BibitemOpen
  \bibfield  {author} {\bibinfo {author} {\bibfnamefont {T.~L.}\ \bibnamefont
  {Gilbert}},\ }\href {\doibase 10.1109/tmag.2004.836740} {\bibfield  {journal}
  {\bibinfo  {journal} {IEEE Trans. Magn.}\ }\textbf {\bibinfo {volume} {40}},\
  \bibinfo {pages} {3443} (\bibinfo {year} {2004})}\BibitemShut {NoStop}%
\bibitem [{SM()}]{SM}%
  \BibitemOpen
  \href@noop {} {\ }\bibinfo {note} {See Supplemental Material for: (1) The
  variation of spin texture and total energy during the iterative process from
  skyrmion to labyrinth domain; (2) The influence of varying supercell sizes on
  the skyrmion spin textures; (3) The magnetic phase diagram of
  antiferromagnetism, LD, skyrmion and ferromagnetism under different Q; (4)
  The relationship of spatial definition and helicity between Néel and Bloch
  skyrmions; (5) The intrinsic properties of CoCl\textsubscript{2}
  monolayer.}\BibitemShut {Stop}%
\bibitem [{\citenamefont {Bessarab}\ \emph {et~al.}(2015)\citenamefont
  {Bessarab}, \citenamefont {Uzdin},\ and\ \citenamefont {Jónsson}}]{RN5280}%
  \BibitemOpen
  \bibfield  {author} {\bibinfo {author} {\bibfnamefont {P.~F.}\ \bibnamefont
  {Bessarab}}, \bibinfo {author} {\bibfnamefont {V.~M.}\ \bibnamefont {Uzdin}},
  \ and\ \bibinfo {author} {\bibfnamefont {H.}~\bibnamefont {Jónsson}},\
  }\href {\doibase 10.1016/j.cpc.2015.07.001} {\bibfield  {journal} {\bibinfo
  {journal} {Comp. Phys. Comm.}\ }\textbf {\bibinfo {volume} {196}},\ \bibinfo
  {pages} {335} (\bibinfo {year} {2015})}\BibitemShut {NoStop}%
\bibitem [{\citenamefont {Bessarab}(2017)}]{RN5278}%
  \BibitemOpen
  \bibfield  {author} {\bibinfo {author} {\bibfnamefont {P.~F.}\ \bibnamefont
  {Bessarab}},\ }\href {\doibase 10.1103/PhysRevB.95.136401} {\bibfield
  {journal} {\bibinfo  {journal} {Phys. Rev. B}\ }\textbf {\bibinfo {volume}
  {95}},\ \bibinfo {pages} {136401} (\bibinfo {year} {2017})}\BibitemShut
  {NoStop}%
\bibitem [{\citenamefont {Kresse}\ and\ \citenamefont
  {Furthm\"uller}(1996)}]{RN70}%
  \BibitemOpen
  \bibfield  {author} {\bibinfo {author} {\bibfnamefont {G.}~\bibnamefont
  {Kresse}}\ and\ \bibinfo {author} {\bibfnamefont {J.}~\bibnamefont
  {Furthm\"uller}},\ }\href {\doibase 10.1103/PhysRevB.54.11169} {\bibfield
  {journal} {\bibinfo  {journal} {Phys. Rev. B}\ }\textbf {\bibinfo {volume}
  {54}},\ \bibinfo {pages} {11169} (\bibinfo {year} {1996})}\BibitemShut
  {NoStop}%
\bibitem [{\citenamefont {Kresse}\ and\ \citenamefont
  {Furthmüller}(1996)}]{RN71}%
  \BibitemOpen
  \bibfield  {author} {\bibinfo {author} {\bibfnamefont {G.}~\bibnamefont
  {Kresse}}\ and\ \bibinfo {author} {\bibfnamefont {J.}~\bibnamefont
  {Furthmüller}},\ }\href {\doibase
  https://doi.org/10.1016/0927-0256(96)00008-0} {\bibfield  {journal} {\bibinfo
   {journal} {Comput. Mater. Sci.}\ }\textbf {\bibinfo {volume} {6}},\ \bibinfo
  {pages} {15} (\bibinfo {year} {1996})}\BibitemShut {NoStop}%
\bibitem [{\citenamefont {Kresse}\ and\ \citenamefont {Hafner}(1993)}]{RN73}%
  \BibitemOpen
  \bibfield  {author} {\bibinfo {author} {\bibfnamefont {G.}~\bibnamefont
  {Kresse}}\ and\ \bibinfo {author} {\bibfnamefont {J.}~\bibnamefont
  {Hafner}},\ }\href {\doibase 10.1103/PhysRevB.47.558} {\bibfield  {journal}
  {\bibinfo  {journal} {Phys. Rev. B}\ }\textbf {\bibinfo {volume} {47}},\
  \bibinfo {pages} {558} (\bibinfo {year} {1993})}\BibitemShut {NoStop}%
\bibitem [{\citenamefont {Kresse}\ and\ \citenamefont {Hafner}(1994)}]{RN72}%
  \BibitemOpen
  \bibfield  {author} {\bibinfo {author} {\bibfnamefont {G.}~\bibnamefont
  {Kresse}}\ and\ \bibinfo {author} {\bibfnamefont {J.}~\bibnamefont
  {Hafner}},\ }\href {\doibase 10.1103/PhysRevB.49.14251} {\bibfield  {journal}
  {\bibinfo  {journal} {Phys. Rev. B}\ }\textbf {\bibinfo {volume} {49}},\
  \bibinfo {pages} {14251} (\bibinfo {year} {1994})}\BibitemShut {NoStop}%
\bibitem [{\citenamefont {Perdew}\ \emph {et~al.}(1996)\citenamefont {Perdew},
  \citenamefont {Burke},\ and\ \citenamefont {Ernzerhof}}]{RN74}%
  \BibitemOpen
  \bibfield  {author} {\bibinfo {author} {\bibfnamefont {J.~P.}\ \bibnamefont
  {Perdew}}, \bibinfo {author} {\bibfnamefont {K.}~\bibnamefont {Burke}}, \
  and\ \bibinfo {author} {\bibfnamefont {M.}~\bibnamefont {Ernzerhof}},\ }\href
  {\doibase 10.1103/PhysRevLett.77.3865} {\bibfield  {journal} {\bibinfo
  {journal} {Phys. Rev. Lett.}\ }\textbf {\bibinfo {volume} {77}},\ \bibinfo
  {pages} {3865} (\bibinfo {year} {1996})}\BibitemShut {NoStop}%
\bibitem [{\citenamefont {Kulish}\ and\ \citenamefont {Huang}(2017)}]{RN145}%
  \BibitemOpen
  \bibfield  {author} {\bibinfo {author} {\bibfnamefont {V.~V.}\ \bibnamefont
  {Kulish}}\ and\ \bibinfo {author} {\bibfnamefont {W.}~\bibnamefont {Huang}},\
  }\href {\doibase 10.1039/C7TC02664A} {\bibfield  {journal} {\bibinfo
  {journal} {J. Mater. Chem. C}\ }\textbf {\bibinfo {volume} {5}},\ \bibinfo
  {pages} {8734} (\bibinfo {year} {2017})}\BibitemShut {NoStop}%
\bibitem [{\citenamefont {Wang}\ \emph {et~al.}(2006)\citenamefont {Wang},
  \citenamefont {Maxisch},\ and\ \citenamefont {Ceder}}]{RN146}%
  \BibitemOpen
  \bibfield  {author} {\bibinfo {author} {\bibfnamefont {L.}~\bibnamefont
  {Wang}}, \bibinfo {author} {\bibfnamefont {T.}~\bibnamefont {Maxisch}}, \
  and\ \bibinfo {author} {\bibfnamefont {G.}~\bibnamefont {Ceder}},\ }\href
  {\doibase 10.1103/PhysRevB.73.195107} {\bibfield  {journal} {\bibinfo
  {journal} {Phys. Rev. B}\ }\textbf {\bibinfo {volume} {73}},\ \bibinfo
  {pages} {195107} (\bibinfo {year} {2006})}\BibitemShut {NoStop}%
\bibitem [{\citenamefont {Botana}\ and\ \citenamefont {Norman}(2019)}]{RN147}%
  \BibitemOpen
  \bibfield  {author} {\bibinfo {author} {\bibfnamefont {A.~S.}\ \bibnamefont
  {Botana}}\ and\ \bibinfo {author} {\bibfnamefont {M.~R.}\ \bibnamefont
  {Norman}},\ }\href {\doibase 10.1103/PhysRevMaterials.3.044001} {\bibfield
  {journal} {\bibinfo  {journal} {Phys. Rev. Mater.}\ }\textbf {\bibinfo
  {volume} {3}},\ \bibinfo {pages} {044001} (\bibinfo {year}
  {2019})}\BibitemShut {NoStop}%
\bibitem [{\citenamefont {Yekta}\ \emph {et~al.}(2021)\citenamefont {Yekta},
  \citenamefont {Hadipour}, \citenamefont {Sasioglu}, \citenamefont
  {Friedrich}, \citenamefont {Jafari}, \citenamefont {Blugel},\ and\
  \citenamefont {Mertig}}]{RN148}%
  \BibitemOpen
  \bibfield  {author} {\bibinfo {author} {\bibfnamefont {Y.}~\bibnamefont
  {Yekta}}, \bibinfo {author} {\bibfnamefont {H.}~\bibnamefont {Hadipour}},
  \bibinfo {author} {\bibfnamefont {E.}~\bibnamefont {Sasioglu}}, \bibinfo
  {author} {\bibfnamefont {C.}~\bibnamefont {Friedrich}}, \bibinfo {author}
  {\bibfnamefont {S.~A.}\ \bibnamefont {Jafari}}, \bibinfo {author}
  {\bibfnamefont {S.}~\bibnamefont {Blugel}}, \ and\ \bibinfo {author}
  {\bibfnamefont {I.}~\bibnamefont {Mertig}},\ }\href {\doibase
  10.1103/PhysRevMaterials.5.034001} {\bibfield  {journal} {\bibinfo  {journal}
  {Phys. Rev. Mater.}\ }\textbf {\bibinfo {volume} {5}},\ \bibinfo {pages}
  {034001} (\bibinfo {year} {2021})}\BibitemShut {NoStop}%
\bibitem [{\citenamefont {Batista}\ \emph {et~al.}(2016)\citenamefont
  {Batista}, \citenamefont {Lin}, \citenamefont {Hayami},\ and\ \citenamefont
  {Kamiya}}]{RN3726}%
  \BibitemOpen
  \bibfield  {author} {\bibinfo {author} {\bibfnamefont {C.~D.}\ \bibnamefont
  {Batista}}, \bibinfo {author} {\bibfnamefont {S.~Z.}\ \bibnamefont {Lin}},
  \bibinfo {author} {\bibfnamefont {S.}~\bibnamefont {Hayami}}, \ and\ \bibinfo
  {author} {\bibfnamefont {Y.}~\bibnamefont {Kamiya}},\ }\href {\doibase
  10.1088/0034-4885/79/8/084504} {\bibfield  {journal} {\bibinfo  {journal}
  {Rep. Prog. Phys.}\ }\textbf {\bibinfo {volume} {79}},\ \bibinfo {pages}
  {084504} (\bibinfo {year} {2016})}\BibitemShut {NoStop}%
\bibitem [{\citenamefont {Hayami}\ \emph {et~al.}(2016)\citenamefont {Hayami},
  \citenamefont {Lin},\ and\ \citenamefont {Batista}}]{RN3724}%
  \BibitemOpen
  \bibfield  {author} {\bibinfo {author} {\bibfnamefont {S.}~\bibnamefont
  {Hayami}}, \bibinfo {author} {\bibfnamefont {S.-Z.}\ \bibnamefont {Lin}}, \
  and\ \bibinfo {author} {\bibfnamefont {C.~D.}\ \bibnamefont {Batista}},\
  }\href {\doibase 10.1103/PhysRevB.93.184413} {\bibfield  {journal} {\bibinfo
  {journal} {Phys. Rev. B}\ }\textbf {\bibinfo {volume} {93}},\ \bibinfo
  {pages} {184413} (\bibinfo {year} {2016})}\BibitemShut {NoStop}%
\bibitem [{\citenamefont {Hayami}(2021)}]{RN2989}%
  \BibitemOpen
  \bibfield  {author} {\bibinfo {author} {\bibfnamefont {S.}~\bibnamefont
  {Hayami}},\ }\href {\doibase 10.1103/PhysRevB.103.224418} {\bibfield
  {journal} {\bibinfo  {journal} {Phys. Rev. B}\ }\textbf {\bibinfo {volume}
  {103}},\ \bibinfo {pages} {224418} (\bibinfo {year} {2021})}\BibitemShut
  {NoStop}%
\bibitem [{\citenamefont {Kharkov}\ \emph {et~al.}(2017)\citenamefont
  {Kharkov}, \citenamefont {Sushkov},\ and\ \citenamefont {Mostovoy}}]{RN3719}%
  \BibitemOpen
  \bibfield  {author} {\bibinfo {author} {\bibfnamefont {Y.~A.}\ \bibnamefont
  {Kharkov}}, \bibinfo {author} {\bibfnamefont {O.~P.}\ \bibnamefont
  {Sushkov}}, \ and\ \bibinfo {author} {\bibfnamefont {M.}~\bibnamefont
  {Mostovoy}},\ }\href {\doibase 10.1103/PhysRevLett.119.207201} {\bibfield
  {journal} {\bibinfo  {journal} {Phys. Rev. Lett.}\ }\textbf {\bibinfo
  {volume} {119}},\ \bibinfo {pages} {207201} (\bibinfo {year}
  {2017})}\BibitemShut {NoStop}%
\bibitem [{\citenamefont {Du}\ \emph {et~al.}(2022)\citenamefont {Du},
  \citenamefont {Dou}, \citenamefont {He}, \citenamefont {Dai}, \citenamefont
  {Huang},\ and\ \citenamefont {Ma}}]{RN3705}%
  \BibitemOpen
  \bibfield  {author} {\bibinfo {author} {\bibfnamefont {W.}~\bibnamefont
  {Du}}, \bibinfo {author} {\bibfnamefont {K.}~\bibnamefont {Dou}}, \bibinfo
  {author} {\bibfnamefont {Z.}~\bibnamefont {He}}, \bibinfo {author}
  {\bibfnamefont {Y.}~\bibnamefont {Dai}}, \bibinfo {author} {\bibfnamefont
  {B.}~\bibnamefont {Huang}}, \ and\ \bibinfo {author} {\bibfnamefont
  {Y.}~\bibnamefont {Ma}},\ }\href {\doibase 10.1021/acs.nanolett.2c00836}
  {\bibfield  {journal} {\bibinfo  {journal} {Nano Lett.}\ }\textbf {\bibinfo
  {volume} {22}},\ \bibinfo {pages} {3440} (\bibinfo {year}
  {2022})}\BibitemShut {NoStop}%
\bibitem [{\citenamefont {Zhong~Shen}\ and\ \citenamefont
  {Song}(2022)}]{RN5231}%
  \BibitemOpen
  \bibfield  {author} {\bibinfo {author} {\bibfnamefont {Z.~W.}\ \bibnamefont
  {Zhong~Shen}, \bibfnamefont {Yufei~Xue}}\ and\ \bibinfo {author}
  {\bibfnamefont {C.}~\bibnamefont {Song}},\ }\href {\doibase
  10.1063/5.0117597} {\bibfield  {journal} {\bibinfo  {journal} {Appl. Phys.
  Lett.}\ }\textbf {\bibinfo {volume} {121}},\ \bibinfo {pages} {202402}
  (\bibinfo {year} {2022})}\BibitemShut {NoStop}%
\bibitem [{\citenamefont {Cui}\ \emph {et~al.}(2021)\citenamefont {Cui},
  \citenamefont {Zhu}, \citenamefont {Jiang}, \citenamefont {Liang},
  \citenamefont {Yu}, \citenamefont {Cui},\ and\ \citenamefont
  {Yang}}]{RN5244}%
  \BibitemOpen
  \bibfield  {author} {\bibinfo {author} {\bibfnamefont {Q.}~\bibnamefont
  {Cui}}, \bibinfo {author} {\bibfnamefont {Y.}~\bibnamefont {Zhu}}, \bibinfo
  {author} {\bibfnamefont {J.}~\bibnamefont {Jiang}}, \bibinfo {author}
  {\bibfnamefont {J.}~\bibnamefont {Liang}}, \bibinfo {author} {\bibfnamefont
  {D.}~\bibnamefont {Yu}}, \bibinfo {author} {\bibfnamefont {P.}~\bibnamefont
  {Cui}}, \ and\ \bibinfo {author} {\bibfnamefont {H.}~\bibnamefont {Yang}},\
  }\href {\doibase 10.1103/PhysRevResearch.3.043011} {\bibfield  {journal}
  {\bibinfo  {journal} {Phys. Rev. Res.}\ }\textbf {\bibinfo {volume} {3}},\
  \bibinfo {pages} {043011} (\bibinfo {year} {2021})}\BibitemShut {NoStop}%
\bibitem [{\citenamefont {Wu}\ \emph {et~al.}(2023)\citenamefont {Wu},
  \citenamefont {Xue}, \citenamefont {Shen},\ and\ \citenamefont
  {Song}}]{RN5241}%
  \BibitemOpen
  \bibfield  {author} {\bibinfo {author} {\bibfnamefont {Z.}~\bibnamefont
  {Wu}}, \bibinfo {author} {\bibfnamefont {Y.}~\bibnamefont {Xue}}, \bibinfo
  {author} {\bibfnamefont {Z.}~\bibnamefont {Shen}}, \ and\ \bibinfo {author}
  {\bibfnamefont {C.}~\bibnamefont {Song}},\ }\href {\doibase
  10.1039/D2CP03860A} {\bibfield  {journal} {\bibinfo  {journal} {Phys. Chem.
  Chem. Phys.}\ }\textbf {\bibinfo {volume} {25}},\ \bibinfo {pages} {96}
  (\bibinfo {year} {2023})}\BibitemShut {NoStop}%
\bibitem [{\citenamefont {Boulle}\ \emph {et~al.}(2016)\citenamefont {Boulle},
  \citenamefont {Vogel}, \citenamefont {Yang}, \citenamefont {Pizzini},
  \citenamefont {de~Souza~Chaves}, \citenamefont {Locatelli}, \citenamefont
  {Mentes}, \citenamefont {Sala}, \citenamefont {Buda-Prejbeanu}, \citenamefont
  {Klein}, \citenamefont {Belmeguenai}, \citenamefont {Roussigne},
  \citenamefont {Stashkevich}, \citenamefont {Cherif}, \citenamefont {Aballe},
  \citenamefont {Foerster}, \citenamefont {Chshiev}, \citenamefont {Auffret},
  \citenamefont {Miron},\ and\ \citenamefont {Gaudin}}]{RN1996}%
  \BibitemOpen
  \bibfield  {author} {\bibinfo {author} {\bibfnamefont {O.}~\bibnamefont
  {Boulle}}, \bibinfo {author} {\bibfnamefont {J.}~\bibnamefont {Vogel}},
  \bibinfo {author} {\bibfnamefont {H.}~\bibnamefont {Yang}}, \bibinfo {author}
  {\bibfnamefont {S.}~\bibnamefont {Pizzini}}, \bibinfo {author} {\bibfnamefont
  {D.}~\bibnamefont {de~Souza~Chaves}}, \bibinfo {author} {\bibfnamefont
  {A.}~\bibnamefont {Locatelli}}, \bibinfo {author} {\bibfnamefont {T.~O.}\
  \bibnamefont {Mentes}}, \bibinfo {author} {\bibfnamefont {A.}~\bibnamefont
  {Sala}}, \bibinfo {author} {\bibfnamefont {L.~D.}\ \bibnamefont
  {Buda-Prejbeanu}}, \bibinfo {author} {\bibfnamefont {O.}~\bibnamefont
  {Klein}}, \bibinfo {author} {\bibfnamefont {M.}~\bibnamefont {Belmeguenai}},
  \bibinfo {author} {\bibfnamefont {Y.}~\bibnamefont {Roussigne}}, \bibinfo
  {author} {\bibfnamefont {A.}~\bibnamefont {Stashkevich}}, \bibinfo {author}
  {\bibfnamefont {S.~M.}\ \bibnamefont {Cherif}}, \bibinfo {author}
  {\bibfnamefont {L.}~\bibnamefont {Aballe}}, \bibinfo {author} {\bibfnamefont
  {M.}~\bibnamefont {Foerster}}, \bibinfo {author} {\bibfnamefont
  {M.}~\bibnamefont {Chshiev}}, \bibinfo {author} {\bibfnamefont
  {S.}~\bibnamefont {Auffret}}, \bibinfo {author} {\bibfnamefont {I.~M.}\
  \bibnamefont {Miron}}, \ and\ \bibinfo {author} {\bibfnamefont
  {G.}~\bibnamefont {Gaudin}},\ }\href {\doibase 10.1038/nnano.2015.315}
  {\bibfield  {journal} {\bibinfo  {journal} {Nat. Nanotechnol.}\ }\textbf
  {\bibinfo {volume} {11}},\ \bibinfo {pages} {449} (\bibinfo {year}
  {2016})}\BibitemShut {NoStop}%
\bibitem [{\citenamefont {Liang}\ \emph {et~al.}(2020)\citenamefont {Liang},
  \citenamefont {Wang}, \citenamefont {Du}, \citenamefont {Hallal},
  \citenamefont {Garcia}, \citenamefont {Chshiev}, \citenamefont {Fert},\ and\
  \citenamefont {Yang}}]{RN3047}%
  \BibitemOpen
  \bibfield  {author} {\bibinfo {author} {\bibfnamefont {J.}~\bibnamefont
  {Liang}}, \bibinfo {author} {\bibfnamefont {W.}~\bibnamefont {Wang}},
  \bibinfo {author} {\bibfnamefont {H.}~\bibnamefont {Du}}, \bibinfo {author}
  {\bibfnamefont {A.}~\bibnamefont {Hallal}}, \bibinfo {author} {\bibfnamefont
  {K.}~\bibnamefont {Garcia}}, \bibinfo {author} {\bibfnamefont
  {M.}~\bibnamefont {Chshiev}}, \bibinfo {author} {\bibfnamefont
  {A.}~\bibnamefont {Fert}}, \ and\ \bibinfo {author} {\bibfnamefont
  {H.}~\bibnamefont {Yang}},\ }\href {\doibase 10.1103/PhysRevB.101.184401}
  {\bibfield  {journal} {\bibinfo  {journal} {Phys. Rev. B}\ }\textbf {\bibinfo
  {volume} {101}},\ \bibinfo {pages} {184401} (\bibinfo {year}
  {2020})}\BibitemShut {NoStop}%
\end{thebibliography}%

\end{document}